\documentclass[aps,prc,preprintnumbers,superscriptaddress,floatfix,showpacs]{revtex4-1}
\usepackage{epsfig}
\usepackage{dcolumn}
\usepackage{bm}
\usepackage{amssymb}

\begin{document}
\title{Suppression of the LHC $p/\pi$ ratio due to the QCD mass spectrum}

\author{Jacquelyn Noronha-Hostler}
\affiliation{Instituto de F\'{i}sica, Universidade de S\~{a}o Paulo, C.P.
66318, 05315-970 S\~{a}o Paulo, SP, Brazil}

\author{Carsten Greiner}
\affiliation{Institut f\"ur Theoretische Physik, Goethe Universit\"at, Frankfurt, Germany}

\date{\today}
\begin{abstract}
Recent measurements of the proton to pion ratio at $\sqrt{s}_{NN}=2.7$ TeV by the ALICE collaboration at the LHC have found it to be lower than predictions from thermal fits.  In this paper we investigate the role that the extended mass spectrum via Hagedorn states- massive resonances that follow an exponential mass spectrum and possess large decay widths- play in the determination of particle ratios at LHC through a scenario of multi-particle reactions and dynamical chemical equilibrium within the hadron gas phase.  We show that it is possible to describe the lower $p/\pi$ ratio at LHC while still obtaining the experimental ratio of $K/\pi$ and $\Lambda/\pi^+$ in the Hagedorn state scenario if the protons are underpopulated at the switching temperature from hydrodynamics to the hadron gas phase. 
\end{abstract}

\maketitle
\section{Introduction}
Particle ratios have been used within thermal models \cite{thermalmodels,StatModel,Schenke:2003mj,RHIC,Andronic:2005yp,Manninen:2008mg,NoronhaHostler:2009tz} to determine the chemical freeze-out temperature and baryon chemical potential at SPS and RHIC \cite{freezeoutline}.  However, recent results from the $\sqrt{s}_{NN}=2.7$ TeV run at ALICE at the LHC \cite{Abelev:2012wca} have measured a proton to pion ratio, $p/\pi=(p+\bar{p})/(\pi^++\pi^-)$ that is lower than predicted by thermal dynamical models \cite{LHC} while other ratios, such as the kaon-to-pion ratio, $K/\pi=(K+\bar{K})/(\pi^++\pi^-)$, and  $\Lambda/\pi^+$ \cite{Abelev:2013xaa}  match expected thermal predictions. It has been theorized that the lower $p/\pi$ ratio can be explained using out of equilibrium final state reactions \cite{Steinheimer:2012rd} calculated within a  hybrid model involving a combination of ideal hydrodynamics \cite{Petersen:2008dd} and UrQMD \cite{urqmd}.  Alternatively, it has been proposed that different hadronization temperatures for light and strange quarks can explain the $p/\pi$ puzzle \cite{Ratti:2011au}.

The UrQMD results in \cite{Steinheimer:2012rd} do not take into account the extended mass spectrum (predicted resonances that have yet to be measured experimentally due to their exponential mass spectrum known as Hagedorn states) or multi-mesonic reactions, yet Hagedorn states and multi-mesonic reactions can also lead to the description of particle ratios at SPS \cite{decaysSPS} and  RHIC \cite{decaysRHIC,Greiner:2004vm,NoronhaHostler:2007jf,NoronhaHostler:2009cf} using dynamical chemical equilibration. Hagedorn states \cite{Hagedorn:1965st} are  resonances, which have yet to be measured experimentall, that follow an exponentially increasing mass spectrum (see \cite{Broniowski:2004yh} for a discussion on the experimental verification of this asymptotic behavior) and have large decay widths.
These Hagedorn states have been found to decrease the shear viscosity to entropy ratio, close to the critical temperature region \cite{NoronhaHostler:2008ju,NoronhaHostler:2012ug,Pal:2010es}. The standard hadron resonance gas (using the know particles from the Particle Data Group), can only match lattice data for the thermodynamical quantities \cite{Borsanyi:2010cj} up to $T\approx 130$ MeV whereas the inclusion of Hagedorn states allows for a description up to $T\approx 155$ MeV \cite{NoronhaHostler:2012ug,Majumder:2010ik}.  Extending the mass spectrum to include these extra resonances has been found to play a role in elliptic flow \cite{Noronha-Hostler:2013ria} and thermal fit models \cite{NoronhaHostler:2009tz}.  The extended mass spectrum can also assist in understanding phase changes from hadronic to deconfined matter and the order of the phase change \cite{Moretto:2005iz,Begun:2009an,Zakout:2006zj,Zakout:2007nb,Ferroni:2008ej,Bugaev:2008iu,Ivanytskyi:2012yx}. It is also postulated that as of yet unmeasured strange baryons predicted from quark models could affect cumulants and correlations of charge fluctuations \cite{Bazavov:2014xya}.    Thus, it is natural to question if they can possibly explain the suppression of the $p/\pi$ ratio at the LHC.  

The lower $p/\pi$ ratio at the LHC, which is unexplained by thermal models,  suggests that assuming that the hadrons being born in chemical equilibrium (i.e. that they freezeout from the quark gluon plasma phase already at their chemical equilibrium values) may not be a good approximation in this case. 
Therefore, it is important to determine the correct mechanism of out-of-equilibrium dynamics that can reproduce this ratio as well as the other particle ratios.  
In this paper we use Hagedorn states within an expanding, cooling fireball and find that through dynamical chemical equilibration we are able to obtain the lower measured $p/\pi$ value when the protons are initially underpopulated at the switching temperature from hydrodynamics to the hadron gas phase.

\section{Model}

The current standard modeling of heavy ion collisions requires some sort of initial conditions followed by relativistic hydrodynamics started a time $t_0$.  At a set switching temperature, $T_{sw}$, the hydrodynamical cells are switched to hadrons using, for instance, the  Cooper Frye approach.  Following this the hadron gas phase is modeled through a hadron transport code such as UrQMD \cite{urqmd} until no more reactions are reached.  In this paper we are interested in determining the correct chemistry within the hadron gas phase that reproduces experimental data.  Thus, for the hydrodynamical modeling we use a simple expanding volume described by a Bjorken expansion  with an accelerating radial flow that begins at $t_0$  described with
\begin{equation}\label{eqn:bjorken}
V(\tau)=\pi\;\tau\left(r_{0}+v_{0}(\tau-\tau_{0})+\frac{1}{2}a_{0}(\tau-\tau_{0})^2 \right)^2
\end{equation}
where $r_0=7.1 \;fm$ is the radius for Pb-Pb and we assume $t_0=0.6$ fm for LHC but also check $\tau_0=1.0$ fm to see the dependence on the initial time.  In most relativistic hydrodynamical codes the initial velocity is set to zero, thus, we take $v_0=0$ and let our acceleration be $a_0=0.03\;fm^{-1}$  (however, we test this assumption in Sec.\ \ref{sec:results}).  We also ensure that our expansion preserves causuality and that the final velocity is reasonable ($v_{final}\approx0.5-0.7$).

The hadron gas phase is modeled using rate equations, which allow for multi-hadron reactions that are catalyzed using Hagedorn states. The evolution of the hadron gas phase (and our rate equations) begin at the switching temperature $T_{sw}$ and end at $T_{end}$.  Considering that in our previous work we found that the Hagedorn state description is adequate to describe lattice data up until $T=155$ MeV, we take our switching temperature to be $T_{sw}=155$ MeV (although we also check the effect of $T_{sw}=165$ MeV). Then, $T_{end}$ is taken as a free parameters- fine tuned according to experimental data, which usually ends up being around $T_{end}\approx135$ MeV. 

Our description of Hagedorn states is taken from \cite{NoronhaHostler:2012ug} wherein we fitted the thermodynamic quantities to recent lattice results \cite{Borsanyi:2010cj}.  In \cite{NoronhaHostler:2012ug} the most relevant Hagedorn spectra were
\begin{eqnarray}
\rho_1&=&A_1 e^{m/T_{H_1}}\label{eqn:rho1}\\
\rho_2&=&\frac{A_2}{\left(m^2+m_{02}^2\right)^{5/4}} e^{m/T_{H_2}}\label{eqn:rho2}\\
\rho_3&=&\frac{A_3}{\left(m^2+m_{03}^2\right)^{3/2}} e^{m/T_{H_3}}\label{eqn:rho3}
\end{eqnarray}
where the parameters are described in Table \ref{tab:par} and are also $\rho_1$, $\rho_2$, and $\rho_3$ in \cite{NoronhaHostler:2012ug}, respectively. Here $T_H$ is the Hagedorn temperature and $A$ is akin to the degeneracy of the Hagedorn states.  We took $A$ and $m_0$ as free parameters, which we use to fit to lattice data. 
\begin{table}
\begin{center}
 \begin{tabular}{|c|c|c|c|c|c|c|}
 \hline
  & $T_H$ (GeV) & $A$  & $m_0$ (GeV)  \\
 \hline
 $\rho_1$ & 0.252 & 2.84 (1/GeV)   &     \\
  $\rho_2$ & 0.180 & 0.63 (GeV$^{3/2}$) & 0.5  \\
 $\rho_3$ & 0.175 & 0.37 (GeV$^{2}$) & 0.5  \\
 \hline
 \end{tabular}
 \end{center}
 \caption{Parameters for the mass spectra shown in Eqs.\ (\ref{eqn:rho1})-(\ref{eqn:rho3}).}
 \label{tab:par}
 \end{table}  
All Hagedorn states must have an exponential term but the prefactor is largely determined by the fit to the mass spectrum.  However, in \cite{Frautschi:1971ij} it was found that the subleading contribution $\approx m^{-a}$ dictates the type of decays that dominate the heavy resonances.  In our case, $\rho_2$ Hagedorn states decay into multi-particle decays,  $\rho_3$ Hagedorn states undergo 2-3 body decays (one Hagedorn state decaying into a lighter Hagedorn states and a known hadron) whereas $\rho_1$ was not considered in \cite{Frautschi:1971ij}. However, as will be shown in Section \ref{sec:results}, $\rho_1$ mimics the behavior of $\rho_2$, which begs the question if they would also favor multi-particle decays.

We use rate equations to describe the particle numbers during the time evolution according to the following decays:
\begin{eqnarray}\label{eqn:decay}
n\pi&\leftrightarrow &HS\leftrightarrow n^{\prime}\pi+X\bar{X}.
\end{eqnarray} 
where $X\bar{X}$ either describes a proton anti-proton pair $p\bar{p}$, a kaon anti-kaon pair $K\bar{K}$, or a lambda anti-lambda pair $\Lambda\bar{\Lambda}$. The rate equations are as follows for the Hagedorn states, pions, and $X\bar{X}$ pair, respectively
\begin{eqnarray}\label{eqn:setpiHSBB}
\dot{N}_{i}&=&\Gamma_{i,\pi}\left[N_{i}^{eq}\sum_{n} B_{i,n}
\left(\frac{N_{\pi}}{N_{\pi}^{eq}}\right)^{n}-N_{i}\right]\nonumber\\
&+&\Gamma_{i,X\bar{X}}\left[ N_{i}^{eq}
\left(\frac{N_{\pi}}{N_{\pi}^{eq}}\right)^{\langle n_{i,x}\rangle} \left(\frac{N_{X\bar{X}}}{N_{X\bar{X}}^{eq}}\right)^2 -N_{i}\right]\nonumber\\
\dot{N}_{\pi }&=&\sum_{i} \Gamma_{i,\pi}  \left[N_{i}\langle n_{i}\rangle-N_{i}^{eq}\sum_{n}
B_{i, n}n\left(\frac{N_{\pi}}{N_{\pi}^{eq}}\right)^{n} \right]\nonumber\\
&+&\sum_{i} \Gamma_{i,X\bar{X}} \langle n_{i,x}\rangle\left[N_{i}-
N_{i}^{eq}
\left(\frac{N_{\pi}}{N_{\pi}^{eq}}\right)^{\langle n_{i,x}\rangle} \left(\frac{N_{X\bar{X}}}{N_{X\bar{X}}^{eq}}\right)^2\right]  \nonumber\\
\dot{N}_{X\bar{X}}&=&\sum_{i}\Gamma_{i,X\bar{X}}\left[ N_{i}- N_{i}^{eq}\left(\frac{N_{\pi}}{N_{\pi}^{eq}}\right)^{\langle n_{i,x}\rangle} \left(\frac{N_{X\bar{X}}}{N_{X\bar{X}}^{eq}}\right)^2\right].
\end{eqnarray}
where the total equilibrium values are $N^{eq}$. We use the grand canonical description to obtain the values of the Hagedorn states at chemical equilibrium. The number density at chemical equilibrium is then
\begin{equation}\label{eqn:neq}
n^{eq}(M_i,T)=\frac{T^3}{2\pi^2}\left[\int_{M_0}^{M_i+\Delta M/2} dm \rho(m)   \int_{0}^{\infty}d\left(\frac{p}{T}\right)\frac{\left(\frac{p}{T}\right)^2}{e^{\sqrt{\left(\frac{p}{T}\right)^2+\left(\frac{m}{T}\right)^2}}-1}-\int_{M_0}^{M_i-\Delta M/2} dm \rho(m)   \int_{0}^{\infty}d\left(\frac{p}{T}\right)\frac{\left(\frac{p}{T}\right)^2}{e^{\sqrt{\left(\frac{p}{T}\right)^2+\left(\frac{m}{T}\right)^2}}-1}\right]
\end{equation}
where $M_i$ is the "mass" of the Hagedorn state, $p$ is the momentum, and $T$ is the temperature.  Because Hageodorn states have a continuous description we must bin the spectrum.  We take the mass $M_i$ in the middle of a mass range of $\Delta M$ and define that as a single Hagedorn state $i$.  We then multiply the number density by the volume to obtain the total equilibrium values i.e. $N^{eq}_i(t)=n^{eq}(M_i,T)V(\tau)$ and the volume is described in Eq.\ (\ref{eqn:bjorken}). Note that we always consider the feed down for the equilibrium values ($n^{eq}$) as described in \cite{NoronhaHostler:2009cf}. We take the minimum mass for a Hagedorn state to be $M_0=1.7$ GeV because above this mass the hadronic mass spectrum begins to deviation from an exponentially increasing behavior.  In this paper we take a maximum mass of $MM=3$ GeV as a very conservative assumption.  However, we all test a higher maximum mass of $MM=5$ GeV in the results below. 

In order to obtain the temperature and time relationship, $T(\tau)$, we assume an isentropic expansion, which then allows us to establish that relationship through solving 
\begin{equation}\label{eqn:temptim}
const=s(T)V(\tau)=\frac{S_\pi}{N_\pi}\int \frac{dN_\pi}{dy}dy
\end{equation}
where $s(T)$ is the entropy at temperature $T$, $\frac{S_\pi}{N_\pi}$ is the average entropy per pion taken as 5.5 \cite{Greiner:1993jn}, and we take the total number of pions at ALICE at LHC is $N_{\pi^+}=733$ and  $N_{\pi^-}=732$ \cite{Abelev:2013vea} for the most $0-5\%$ central collisions whereas for $N_{\pi^0}$ we simply take the average and, therefore, arrive at $N_{\pi^{all}}=\int \frac{dN_\pi}{dy}dy=2197.5$. 

The decay widths, 
\begin{equation}\label{eqn:decaywidths}
\Gamma_i [GeV]=0.15m_i-0.027
\end{equation}
are found by fitting the known decays widths of the non-strange, mesonic particle to a linear fit for further explanation see \cite{NoronhaHostler:2007jf,NoronhaHostler:2009cf} and this is similar to what was done in \cite{Lizzi:1990na,Lizzi:1990if}.  It follows then that we separate the decay width into the pion and $X\bar{X}$ contribution $\Gamma_{i}=\Gamma_{i,\pi}+\Gamma_{i,X\bar{X}}$ where $\Gamma_{i,X\bar{X}}=\langle X\rangle_i \Gamma_i$ and $\langle X\rangle_i$ is the average number of $X$ that the $i^{th}$ Hagedorn state decays into. Our values for $\langle X\rangle_i$ are taken from a microcanonical model \cite{Liu,Greiner:2004vm,NoronhaHostler:2009cf} where further details are given in \cite{NoronhaHostler:2009cf}. 

The branching ratios $B_{i,n}$ for the decay $HS\leftrightarrow n\pi$ and $\langle n_{i,x}\rangle$ for the decay $HS\leftrightarrow n^{\prime}\pi+X\bar{X}$ are obtained from the aforementioned microcanonical model \cite{Liu,Greiner:2004vm}.  However, in \cite{Liu,Greiner:2004vm} the average number of pions was only calculated for $HS\leftrightarrow n^{\prime}\pi+p\bar{p}$, thus, here we assume that for the production of a kaon anti-kaon pair that approximately double the number of pions because kaons are roughly half the mass of pions and for lambda we assume $\langle n_{i,p}\rangle=\langle n_{i,\Lambda}\rangle$ since the mass of the proton and lambda particle are not significantly different.

\section{Results}\label{sec:results}

As we discussed in \cite{NoronhaHostler:2012ug}, the description of the Hagedorn state mass spectrum is relatively robust when it comes to fitting thermodynamical quantities.  Because of that it is difficult to eliminate one specific type of description of the extended mass spectrum.  Therefore, in this paper we investigate if it is possible to more precisely ascertain the Hagedorn mass spectrum description using the particle ratios.  

\begin{figure}
\centering
\includegraphics[width=4.in]{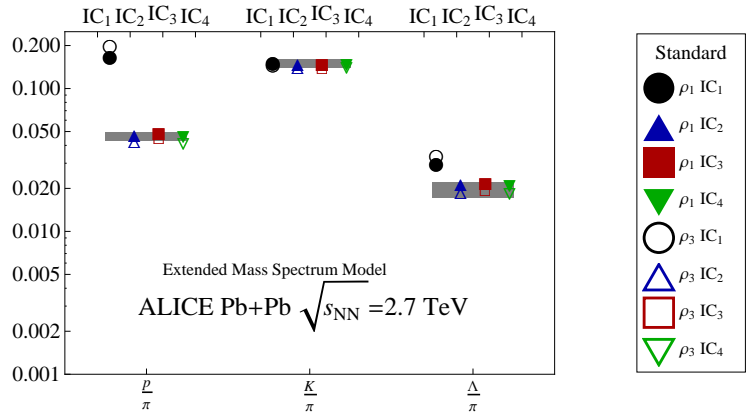} 
\caption{Results for  $p/\pi=(p+\bar{p})/(\pi^++\pi^-)$,  $K/\pi=(K+\bar{K})/(\pi^++\pi^-)$, and $\Lambda/\pi^+$ for $\rho_1$ ($T_{end}=133$ MeV) and $\rho_3$ ($T_{end}=128$ MeV), using $\tau_0=0.6 fm$, and  $T_{sw}=155$ MeV for various initial conditions (see Table \ref{tab:IC}). $\rho_2$ is not shown because the results are almost identical to those shown in $\rho_1$.} \label{fig:standard}
\end{figure}

Initially, we take our best guess for the parameters taken from current assumptions used in hydrodynamical calculations and other models.  In Fig.\ \ref{fig:standard} we show the results for the final ratios of $p/\pi=(p+\bar{p})/(\pi^++\pi^-)=0.046\pm0.003$,  $K/\pi=(K+\bar{K})/(\pi^++\pi^-)=0.149\pm0.01$, and $\Lambda/\pi^+=0.0194\pm0.0025$ for only $\rho_1$ and $\rho_3$, $\tau_0=0.6$ fm, and $T_{sw}=155$ MeV.  After comparing to experimental values we found $T_{end}=133$ MeV for $\rho_1$, $T_{end}=136$ MeV for  $\rho_2$, and $T_{end}=128$ MeV for $\rho_3$.   The reason for only showing two $\rho$'s in Fig.\ \ref{fig:standard} is because the results for $\rho_2$ are almost the same as $\rho_1$ and the variables would clutter the graph without warrant.  In the remaining graphs all three $\rho$'s will be shown. The various initial conditions (i.e. the initial number of $\pi$'s, $X\bar{X}$ pairs's, and Hagedorn states) are defined in Table \ref{tab:IC}.  
\begin{table}
\begin{center}
 \begin{tabular}{|c|c|c|c|}
 \hline
 & & & \\
   & $\frac{N_{\pi}}{N_{\pi}^{eq}}(\tau_0)$ & $\frac{N_{i}}{N_{i}^{eq}}(\tau_0)$ & $\frac{N_{X\bar{X}}}{N_{X\bar{X}}^{eq}}(\tau_0)$ \\
    & & & \\
 \hline
 $IC_1$ & 1 & 1 & 1 \\
$IC_2$ & 1 & 1 & 0 \\
$IC_3$ & 1.1 & 0.5 & 0 \\
$IC_4$ & 0.95 & 1.2 & 0 \\
 \hline
 \end{tabular}
 \end{center}
 \caption{Initial condition configurations.}\label{tab:IC}
 \end{table}

As you can see in Fig.\ \ref{fig:standard}, there appears to be a dependence on our description of Hagedorn states.  Comparing the description of the mass spectrum, $\rho_3$  provides slightly lower values for the various ratios whereas $\rho_1$ (and $\rho_2$) are slightly higher.  All $\rho$'s managed to adequately reproduce the ALICE ratios when there are only a small number of initial protons in the system and any variation on the initial number of pions or Hagedorn states. On the other hand, if the protons, pions, and Hagedorn States all begin in equilibrium then $p/\pi$ ratio is too large (for all $\rho$'s) and also for the $\Lambda/\pi⁺$ ratio.

Because the initial conditions do not strongly affect the final ratios for small variations of the initial number of pions and Hagedorn states (compare $IC_2$, $IC_3$, and $IC_4$), in the following, we test using only two initial conditions: $IC_1$ where all the hadrons begin in chemical equilibrium and $IC_2$ where only the pions and Hagedorn states begin in chemical equilibrium and the rest begin at zero. We expect (if our parameters are robust) that $IC_1$ will end up with too high of a $p/\pi$ regardless of our change in parameters.  $IC_2$, on the other hand, is an extreme case where there are no initial protons or pions present.  Realistically, there will most likely be at least some protons, kaons, and lambdas present in each event.  However, the results for those ratios should then fall in between $IC_1$ at the high extreme and $IC_2$ at the lower extreme.  


\subsection{Initial Time and Expansion}
\begin{figure}
\centering
\includegraphics[width=3.in]{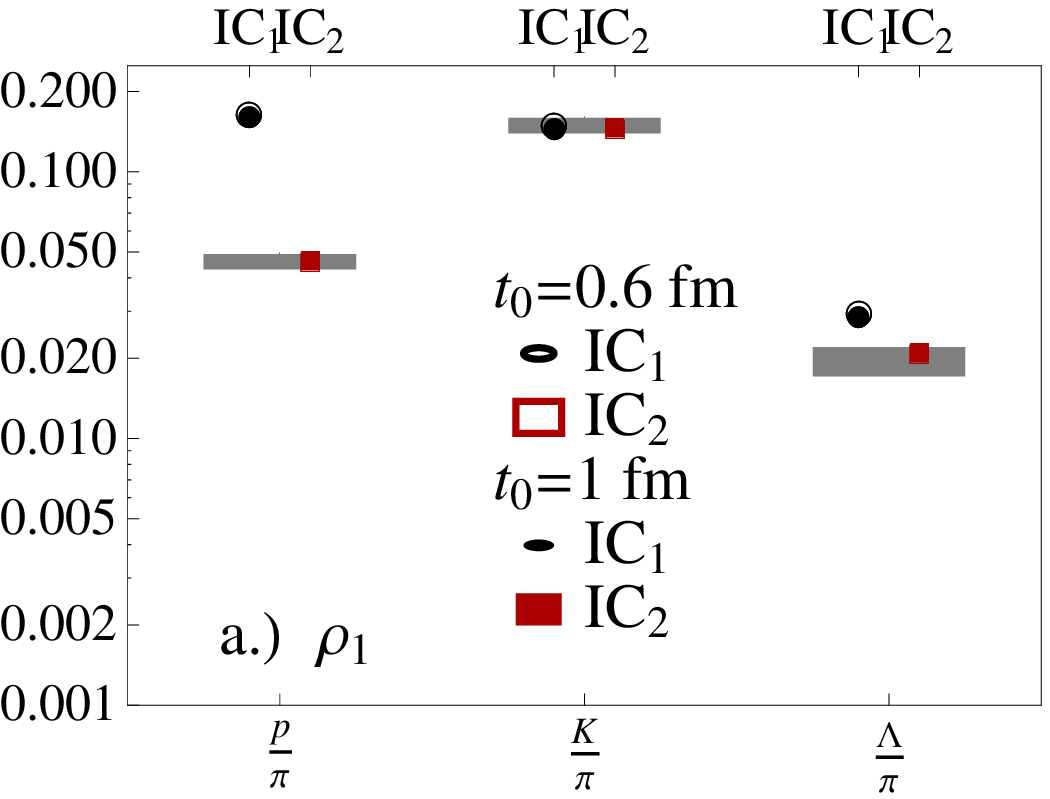}  \\
\includegraphics[width=3.in]{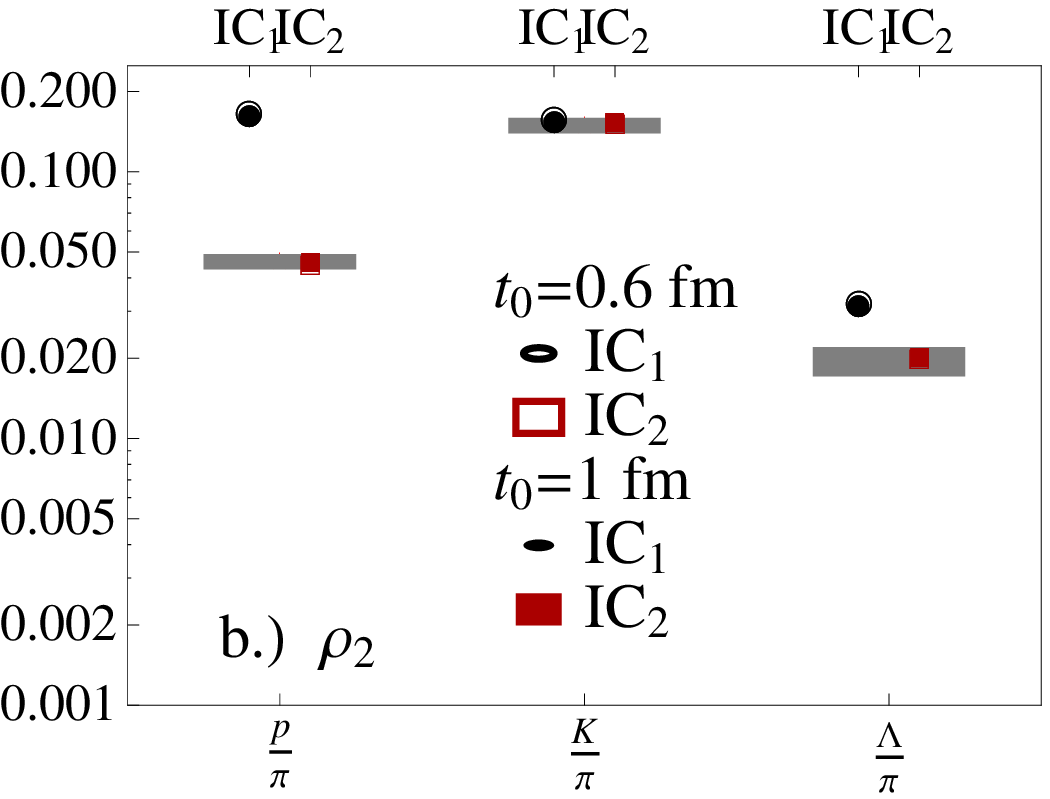} \\
\includegraphics[width=3.in]{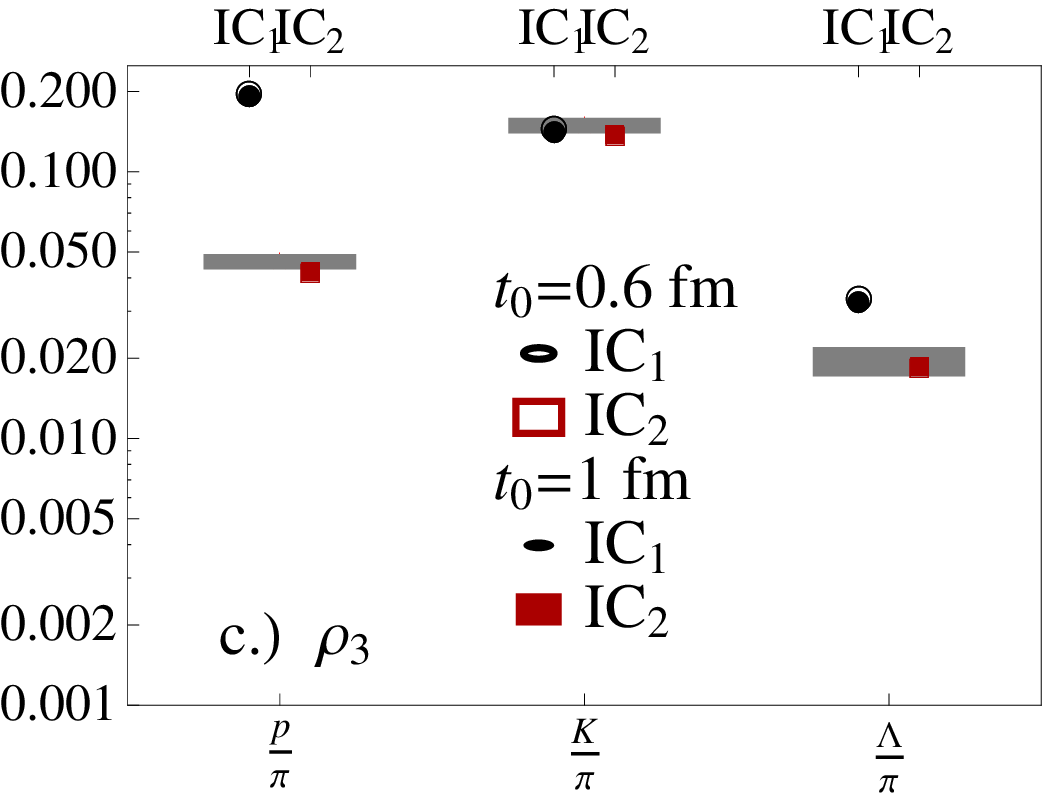} 
\caption{Results for  $p/\pi=(p+\bar{p})/(\pi^++\pi^-)$,  $K/\pi=(K+\bar{K})/(\pi^++\pi^-)$, and $\Lambda/\pi^+$ for a.) $\rho_1$, b.) $\rho_2$, c.) $\rho_2$, using  $T_{sw}=155$ MeV while varying between $\tau_0=0.6 fm$ (solid shapes) and $\tau_0=1.0 fm$ (outlined shapes).} \label{fig:t01}
\end{figure}

We use Bjorken expansion  with an accelerating radial flow to describe the expansion of our system (see Eq.\ (\ref{eqn:bjorken}) ), which requires the input of the constants $\tau_0$, $a_0$, $v_0$, and $r_0$.  We set $v_0=0$ following the example of the many initial conditions for relativistic hydrodynamical codes and let $r_0=7.1$ fm be the radius of a gold atom. This leaves us with the remaining $\tau_0$ and $a_0$ to test their robustness.

Usually, one assumes that at LHC relativistic hydrodynamics begins at $\tau_0=0.6$ fm and that at RHIC it begins at $\tau_0=1$ fm.  Here we varied the initial time using both  $\tau_0=0.6$ fm and $\tau_0=1$ fm.  In Fig.\ \ref{fig:t01} the difference between $\tau_0=0.6$ fm and $\tau_0=1$ fm are the difference between the outlined points ($\tau_0=0.6$ fm) and solid points ($\tau_0=1$ fm).  Regardless of the initial conditions there is little change between the different $\tau_0$'s.  However, the amount of time spent within the hadron gas phase is slightly larger than for $\tau_0=0.6$ fm, while the final temperature $T_{end}$ is identical to $\tau_0=0.6$ fm.

\begin{figure}
\centering
\includegraphics[width=3.in]{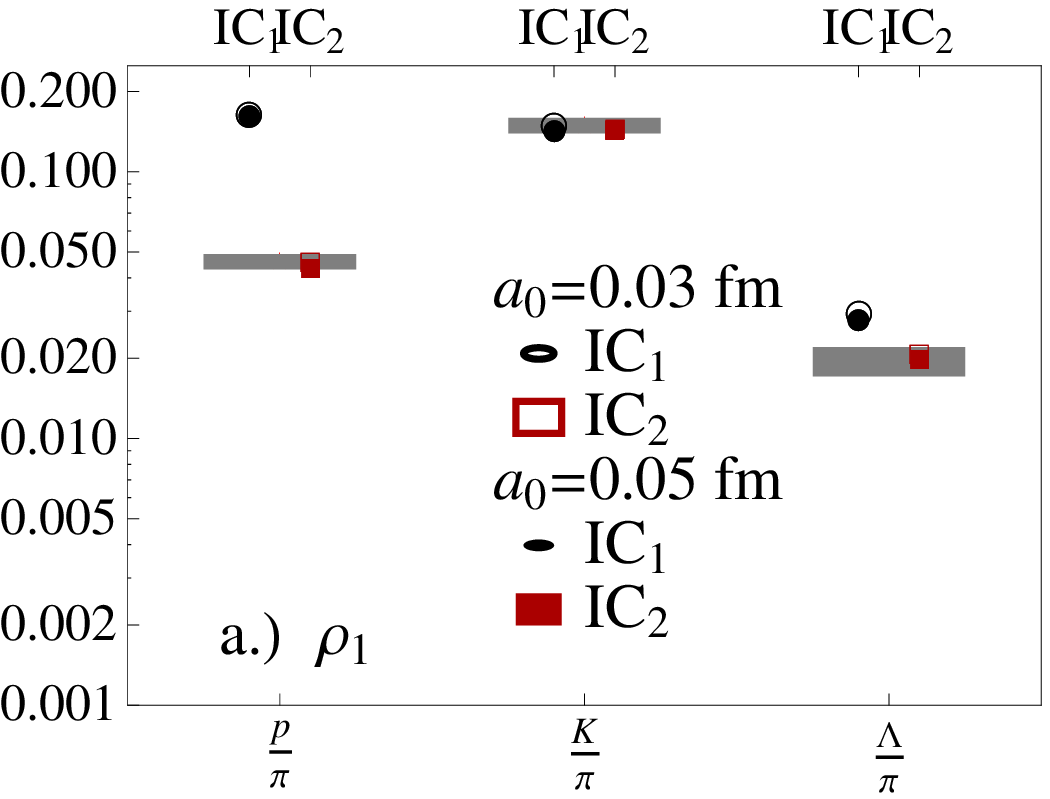}  \\
\includegraphics[width=3.in]{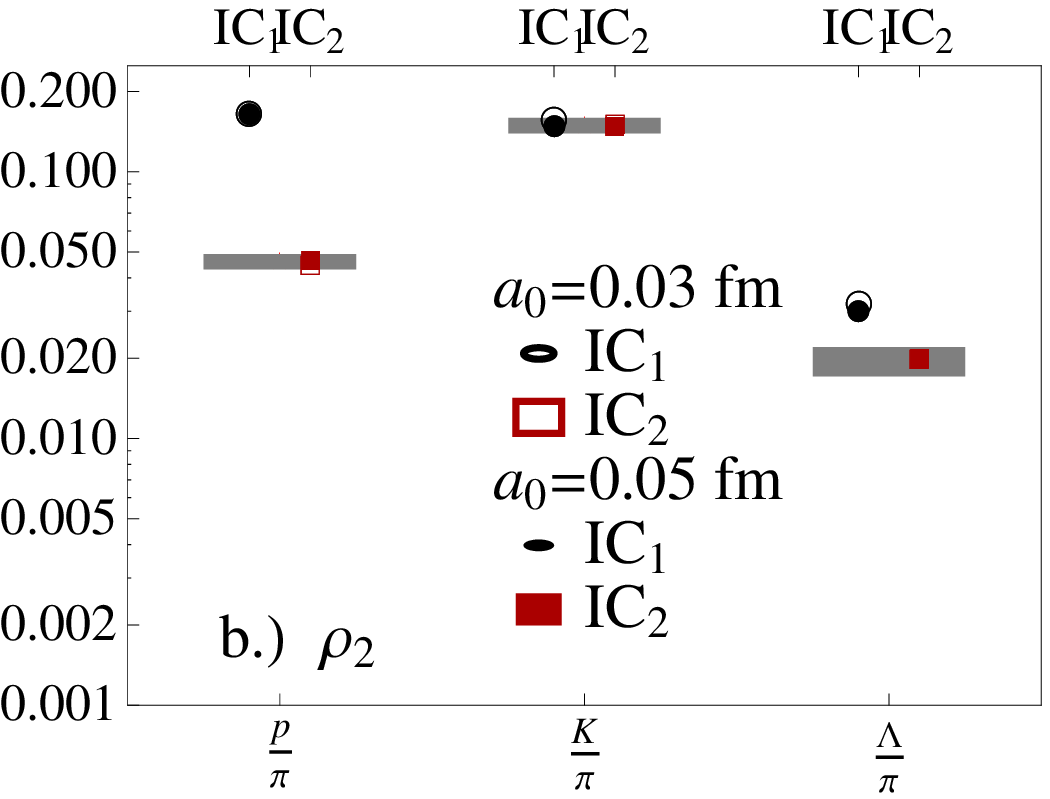} \\
\includegraphics[width=3.in]{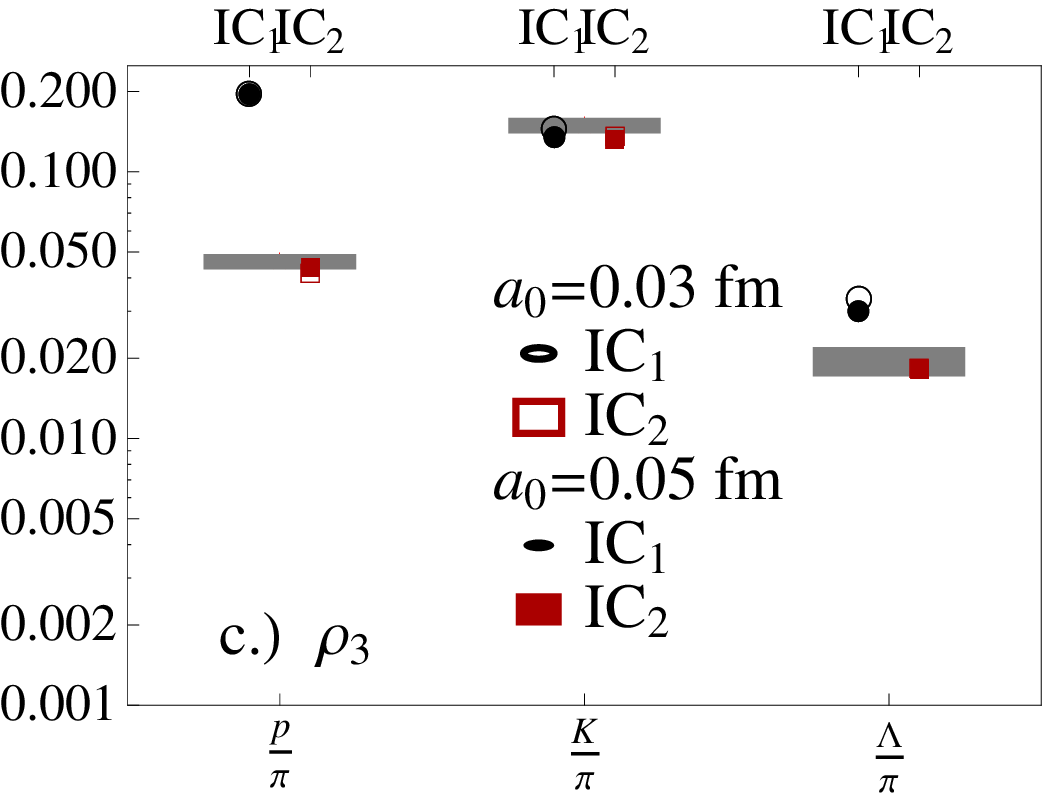} 
\caption{Results for  $p/\pi=(p+\bar{p})/(\pi^++\pi^-)$,  $K/\pi=(K+\bar{K})/(\pi^++\pi^-)$, and $\Lambda/\pi^+$ for a.) $\rho_1$, b.) $\rho_2$, c.) $\rho_2$, using  $T_{sw}=155$ MeV and $\tau_0=0.6 fm$ while varying between $a_0=0.03 fm$ (solid shapes) and $a_0=0.05 fm$ (outlined shapes).} \label{fig:a01}
\end{figure}

However, when one compares the effect of the accelerations, $a_0$, we find that the amount of time spent in the system is significantly shorter for a larger acceleration (not surprising) but a lower $T_{end}$ is needed.  In Fig. \ref{fig:a01} one can clearly see that there is almost no affect on the end results regardless of initial conditions or mass spectrum description when we increase the acceleration from $a_0=0.03 fm$ to $a_0=0.05 fm$ (especially for the case when there are no initial population of the $X\bar{X}$ pairs).  The expansion for $a_0=0.05 fm$ takes roughly $1$ fm less time to reach the experimental values, however,  $T_{end}$ lowers roughly 3-5 MeV (as in $T_{end}=136$ MeV for $\rho_2$ and  $a_0=0.01 fm$ but drops to  $T_{end}=132$ MeV with the increased acceleration). 

\subsection{Switching Temperature}

\begin{figure}
\centering
\includegraphics[width=3.in]{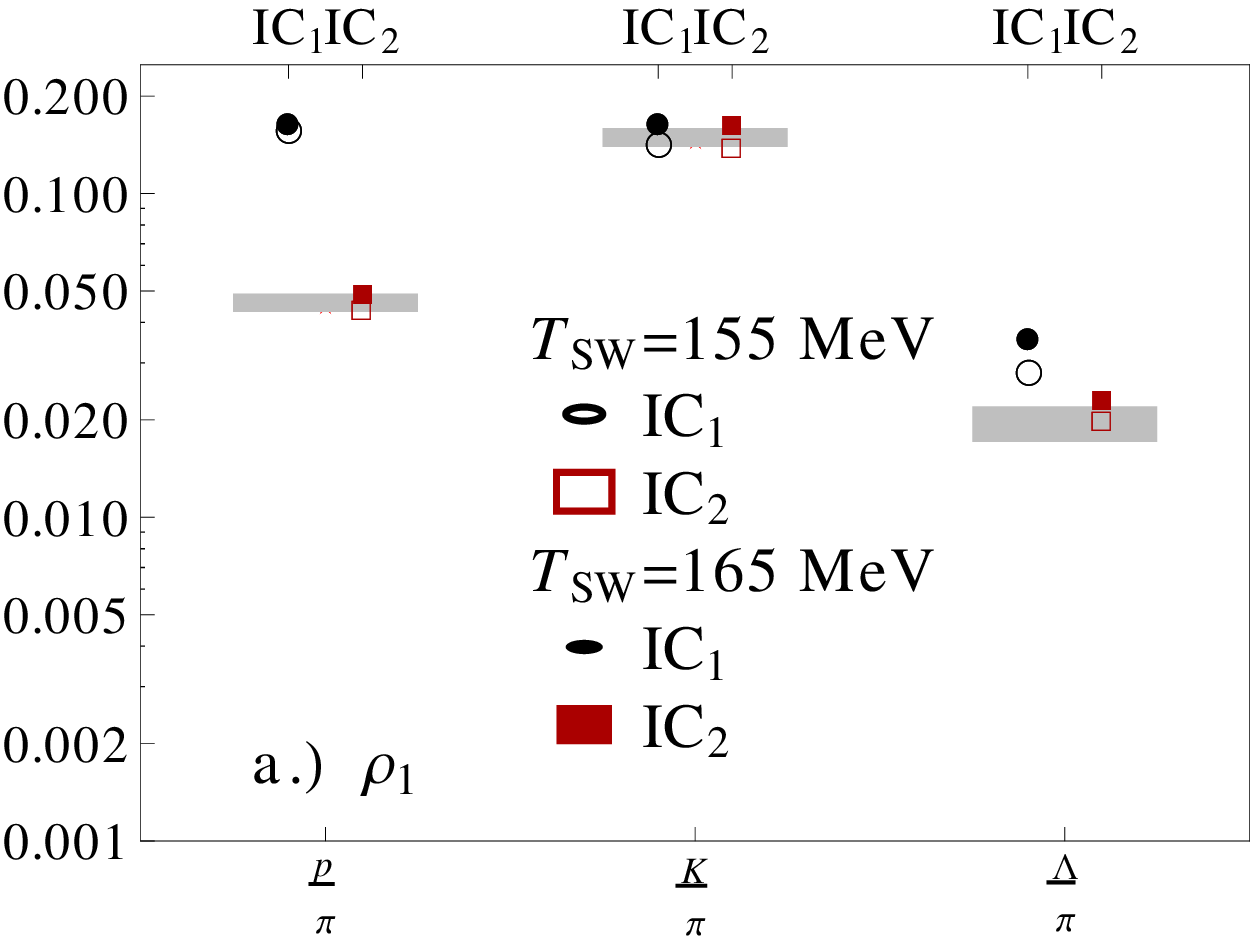}  \\
\includegraphics[width=3.in]{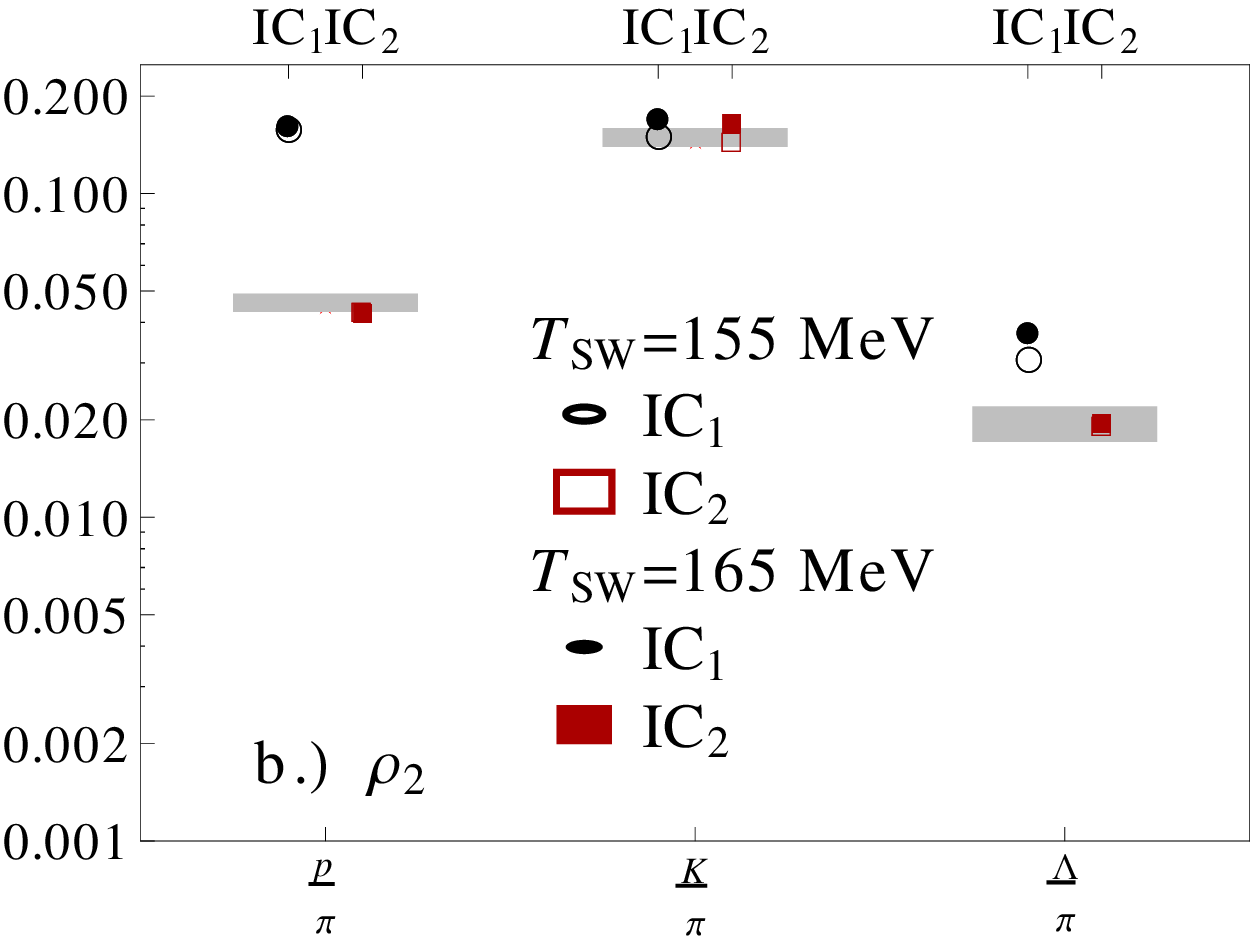} \\
\includegraphics[width=3.in]{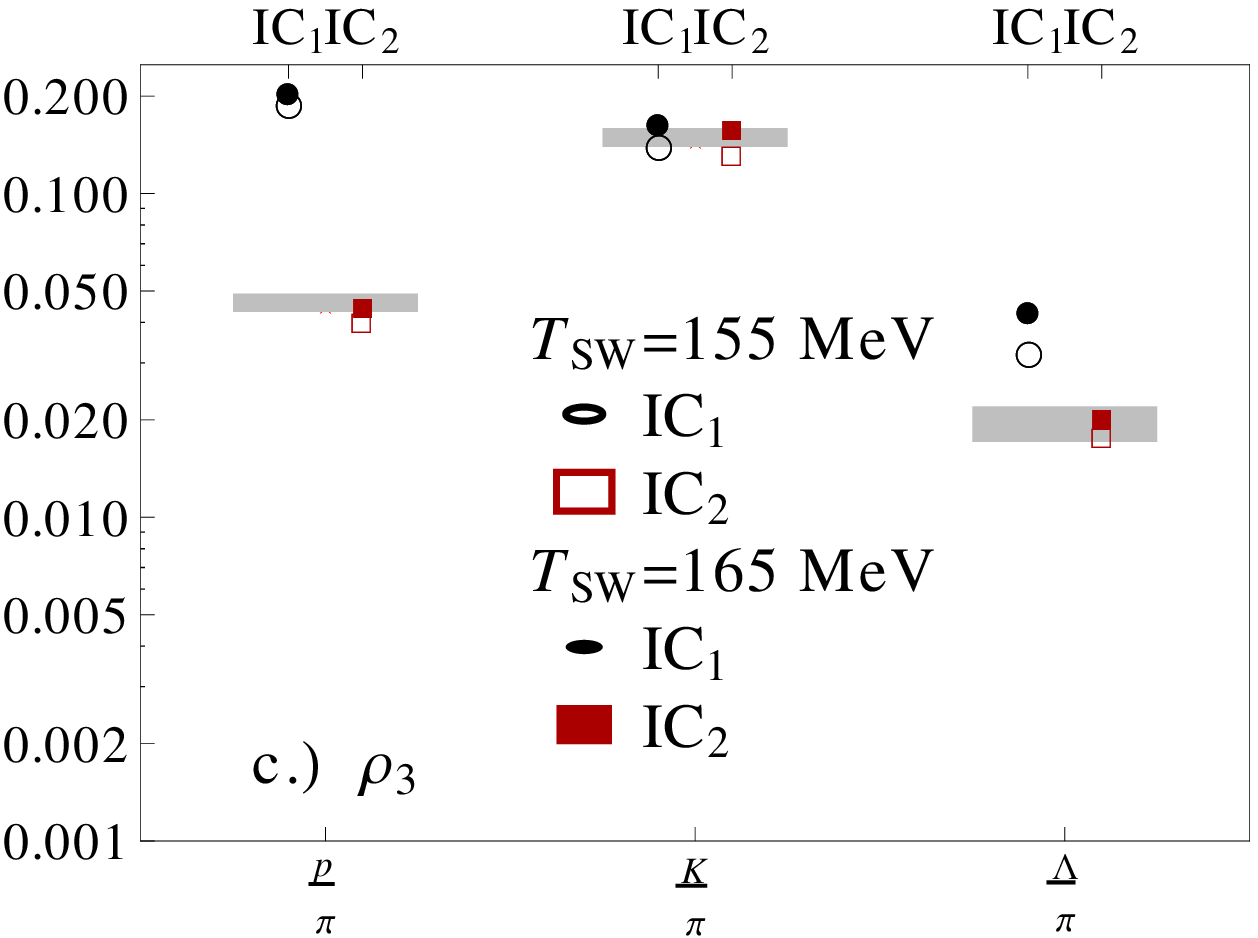} 
\caption{Results for  $p/\pi=(p+\bar{p})/(\pi^++\pi^-)$,  $K/\pi=(K+\bar{K})/(\pi^++\pi^-)$, and $\Lambda/\pi^+$ for  a.) $\rho_1$, b.) $\rho_2$, c.) $\rho_2$, using $\tau_0=0.6 fm$ where we varied the switching temperature between $T_{sw}=155$ MeV  and $T_{sw}=165$ MeV.} \label{fig:Tsw}
\end{figure}

The switching temperature $T_{sw}$ is the temperature where we begin the hadron decays following a hydrodynamical expansion.  In \cite{Majumder:2010ik,NoronhaHostler:2012ug} it was shown that the hadron resonance gas can only describe the thermodynamical quantities calculated from the lattice up until about $T\approx 130-140$ MeV whereas the implementation of Hagedorn states can increase the described equation of state up until about $T=155$ MeV.  Furthermore, we found in  \cite{Noronha-Hostler:2013ria} that the elliptical flow is dependent on the inclusion of Hagedorn states, especially if one uses a higher switching temperature.  Therefore, it is important to check the effects of decays including Hagedorn states at various switching temperatures.

In Fig.\ \ref{fig:Tsw} we test a higher switching temperature of $T_{sw}=165$ MeV against our usual one of $T_{sw}=155$ MeV and find that in general the ratios are somewhat larger than for a lower switching temperature.  Especially for $\rho_1$ the ratios are all at the very high end of the error bars or even slightly above them (for $\Lambda/\pi$).  While the difference is small these results do indicate that one must be careful when considering a higher switching temperature. Only for $\rho_3$ are the adjusted particle ratios slighly improved because the standard $\rho_3$ particle ratios seen in Fig.\ \ref{fig:standard} are on the lower end of the experimental data. It is also not surprising that more difficulties matching the particle ratios are seen for a higher switching temperature considering that in this region the thermodynamical quantities no longer match lattice data. One should note that a higher switching temperature also corresponds to a significantly higher ending temperature where $T_{end}=150$, 154, and 147, respectively.  

\subsection{Decay Widths}

\begin{figure}
\centering
\includegraphics[width=3.in]{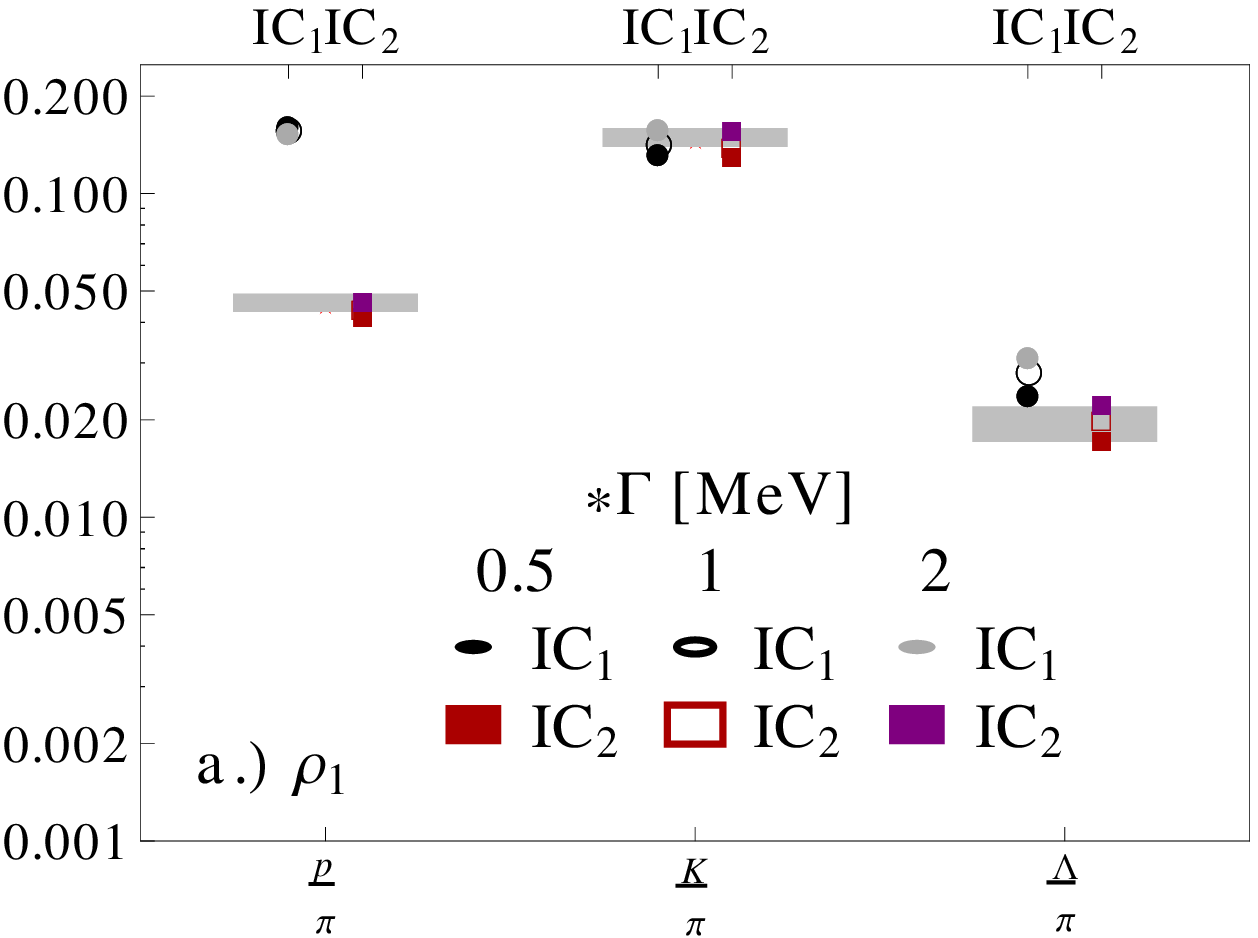}  \\
\includegraphics[width=3.in]{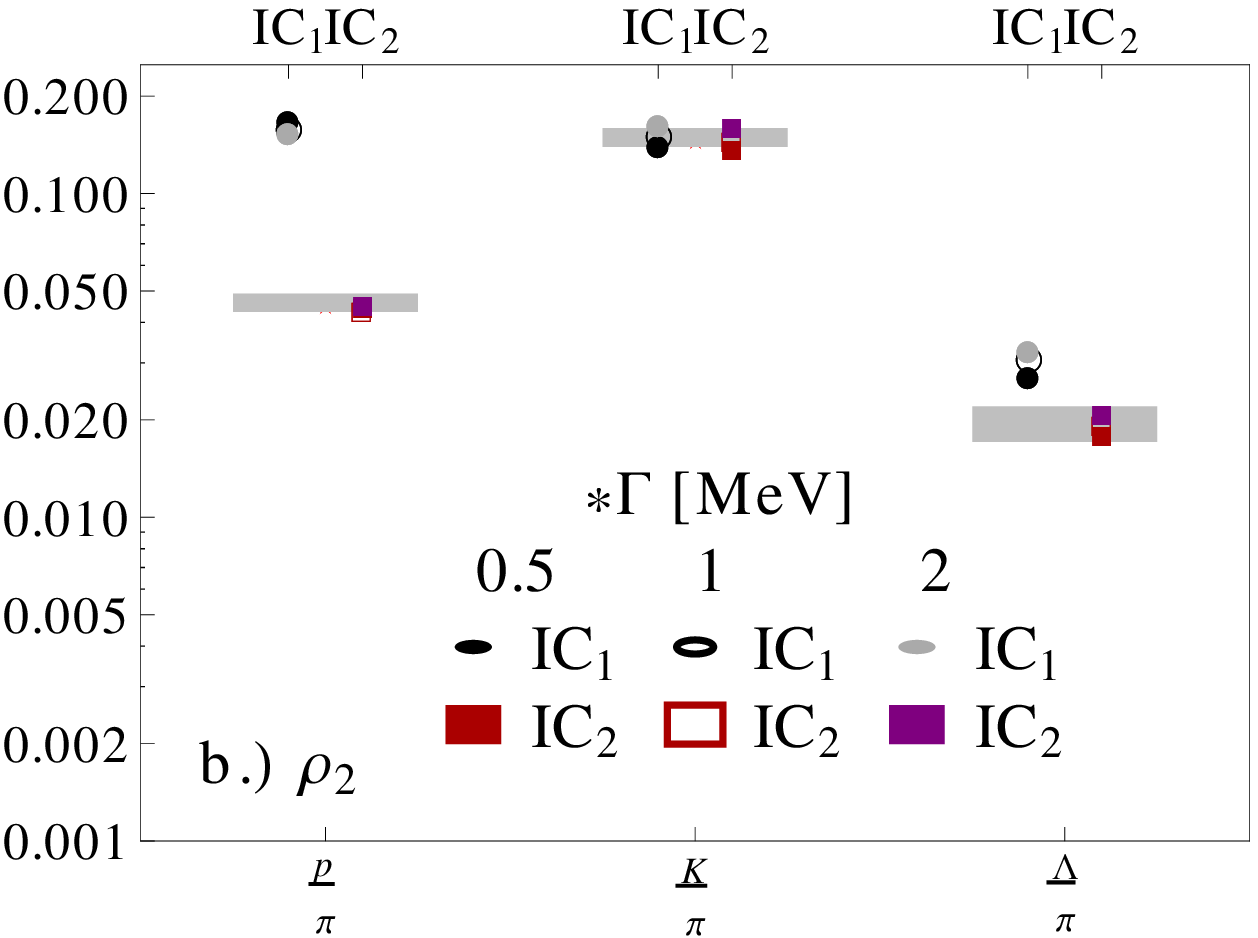} \\
\includegraphics[width=3.in]{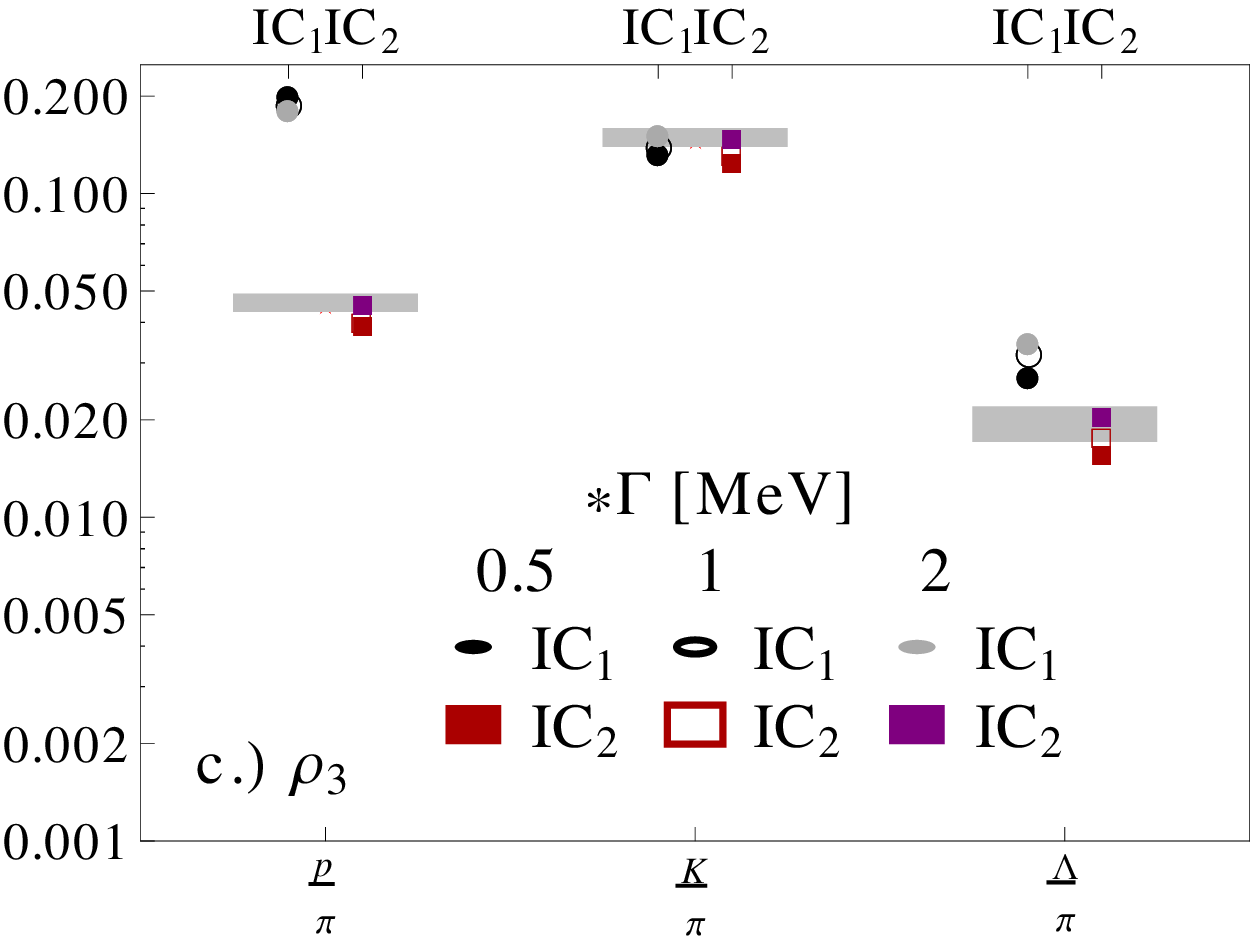} 
\caption{Results for  $p/\pi=(p+\bar{p})/(\pi^++\pi^-)$,  $K/\pi=(K+\bar{K})/(\pi^++\pi^-)$, and $\Lambda/\pi^+$ for  a.) $\rho_1$, b.) $\rho_2$, c.) $\rho_2$, using $\tau_0=0.6 fm$ and  $T_{sw}=155$ MeV while varying the decay withs $\Gamma_i(M_i)$.} \label{fig:gam}
\end{figure}

Hagedorn states are resonances that have yet to be measured, so we have no experimental data to determine their decay widths.  In this paper we assume that there is a linear increase with the mass of the Hagedorn states that we fitted using the known resonancs. However, it is not unrealistic to question that assumption.  Additionally, there are other models \cite{Beitel:2014kza,Pal:2005rb} that describe the decay width and cross-section of Hagedorn states, although they all generally find that the Hagedorn states provide reasonable particle ratios for  experimental measurements. Thus, in this section we check the robustness of our assumption by taking the following variations of our initial decay widths in Eq.\ (\ref{eqn:decaywidths}): $0.5\Gamma_i$, $\Gamma_i$, and $2\Gamma_i$.

It appears that $\rho_1$ and $\rho_3$ are more sensitive to the size of the decay width than $\rho_2$.  For $\rho_1$ and $\rho_3$ the $\Lambda/\pi^+$ is especially sensitive to our choice in the decay width where it is clearly below experimental results for $0.5\Gamma_i$.  On the other hand, $0.5\Gamma_i$ results are also on the low side for the $K/\pi$ (although for $\rho_2$ they manage to fit within the error bars). When we increase our decay width to $2\Gamma_i$ we see that the ratios increase, however, they still remain within the error bars for all $\rho$'s and particle ratios. This indicates that the decay width could be increased without much affect on the comparison to data, however, there appears to be a lower limit that you can decrease the decay width before it has troubles fitting the $\Lambda/\pi^+$ and $K/\pi$  ratio. 

In the case of $\rho_3$ it appears that the increased decay width of $2\Gamma_i$ provides a better fit across the board whereas decreasing the decay width universally worsens the ratios.  In all cases the initial condition of all hadrons being born in chemical equilibrium provides a too high $p/\pi$ regardless of the decay width.  Finally, we find that the decay width has a strong impact on $T_{end}$ where $0.5\Gamma_i$ lowers $T_{end}\approx 120$ MeV and $2\Gamma_i$ increases $T_{end}\approx 140$ MeV.

\subsection{Maximum Mass}

\begin{figure}
\centering
\includegraphics[width=3.in]{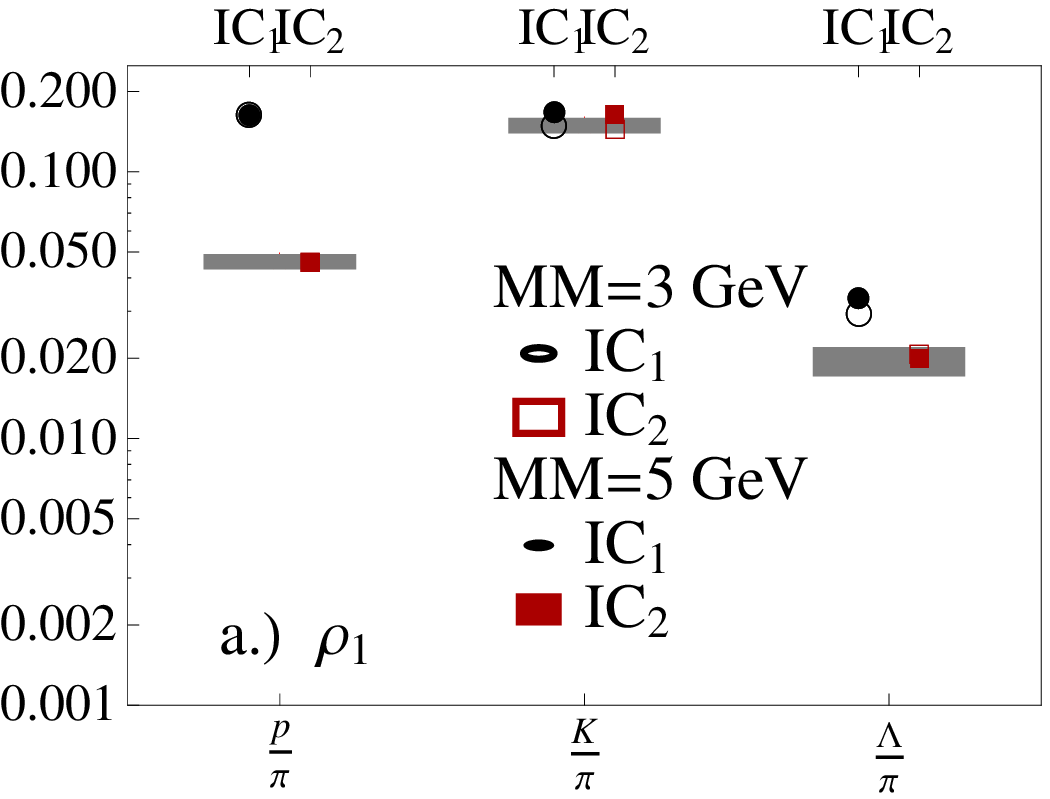}  \\
\includegraphics[width=3.in]{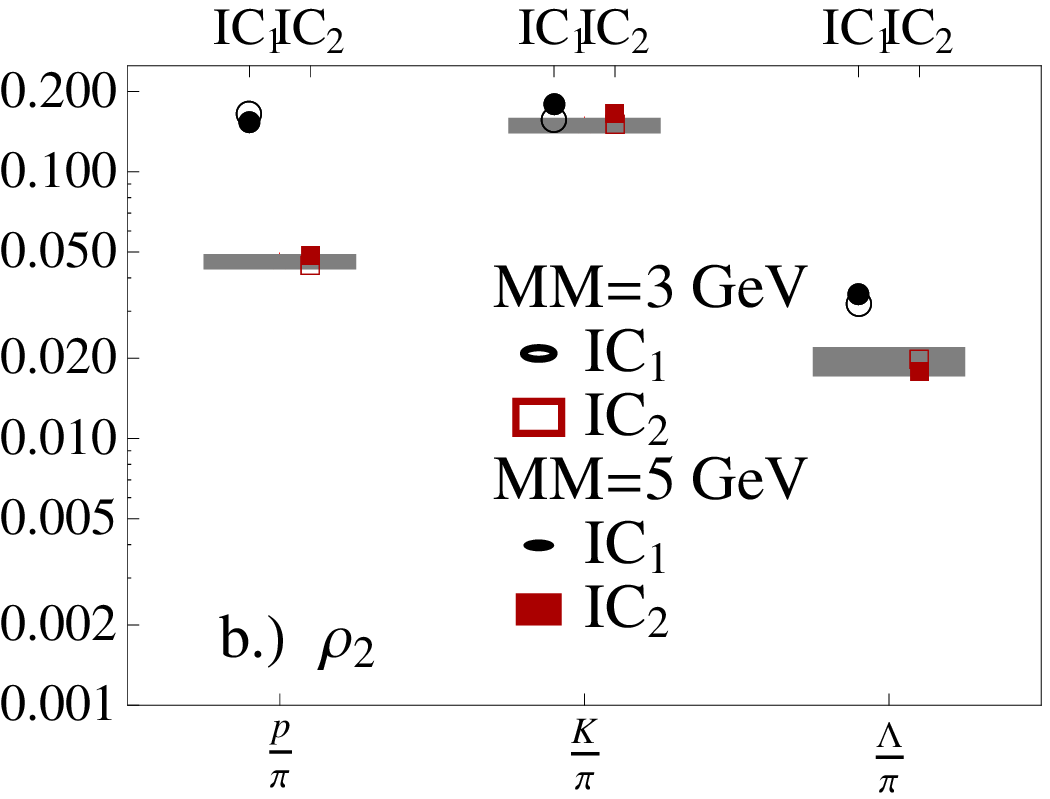} \\
\includegraphics[width=3.in]{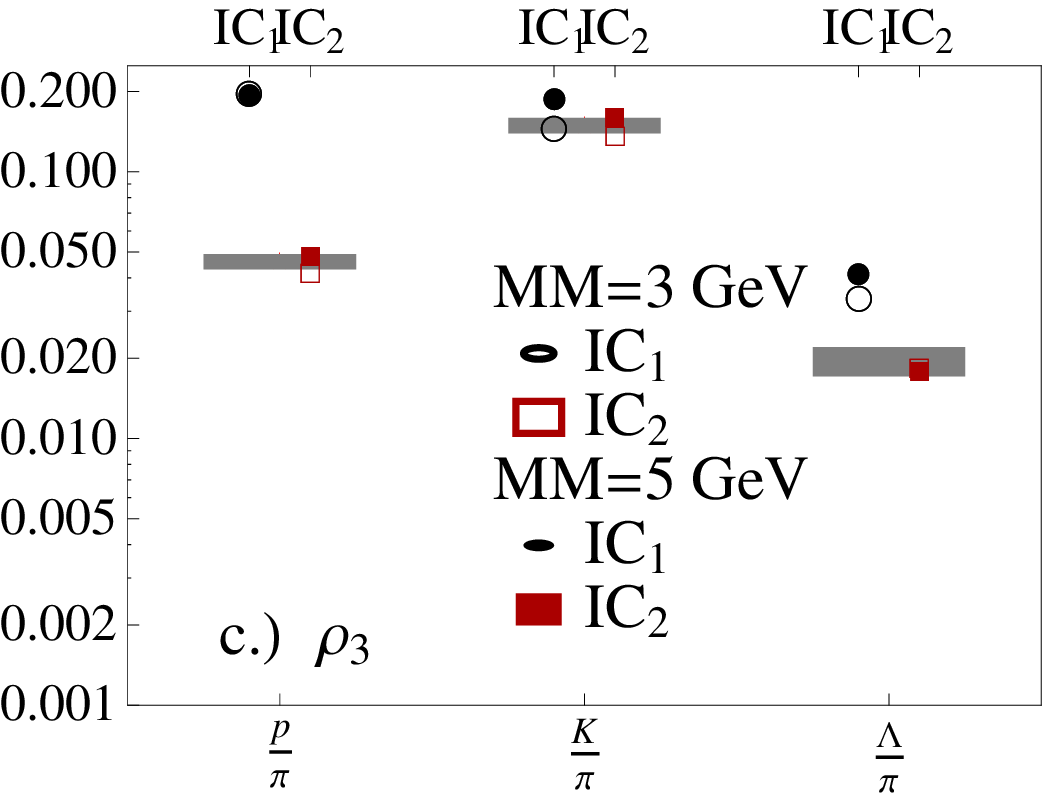} 
\caption{Results for $p/\pi=(p+\bar{p})/(\pi^++\pi^-)$,  $K/\pi=(K+\bar{K})/(\pi^++\pi^-)$, and $\Lambda/\pi^+$ for  a.) $\rho_1$, b.) $\rho_2$, c.) $\rho_2$, using $\tau_0=0.6 fm$ and $T_{sw}=155$ MeV while varying the maximum mass of the Hagedorn States between $MM=3$ GeV and $MM=5$ GeV.} \label{fig:MM}
\end{figure}

In principle, the Hagedorn mass spectrum should be extended up to infinity.  In practice, we need to set a cutoff for the upper mass.  Realistically when it comes to measuring Hagedorn states experimentally, it becomes increasingly more difficult to measure larger, heavier resonances due to their large decay widths such that they decay almost immediately.  Thus, in this paper we chose to take a more conservative approach and only include Hagedorn states up to $MM=3$ GeV.  

In Fig.\ \ref{fig:MM} we compare the results for $MM=3$ GeV and $MM=5$ GeV and find that for all mass spectra that the increase in maximum mass primarily increases the $K/\pi$ while also slightly increases the $p/\pi$ for $\rho_2$ and $\rho_3$ (while still staying within the experimental error bars).  For all the $\rho$'s we find that the time of the expansion is roughly $2$ fm shorter and that $T_{end}$ is higher at $T_{end}^{\rho_1}=140$ MeV, $T_{end}^{\rho_2}=146$ MeV and $T_{end}^{\rho_3}=140$ MeV.  This implies that a larger maximum mass correlates to a higher chemical freeze-out temperature, however, it does not appear to affect the fit to the experimental values much. 



\section{Detailed Balance and the Total Number of Protons}

\begin{figure}
\includegraphics[width=3.in]{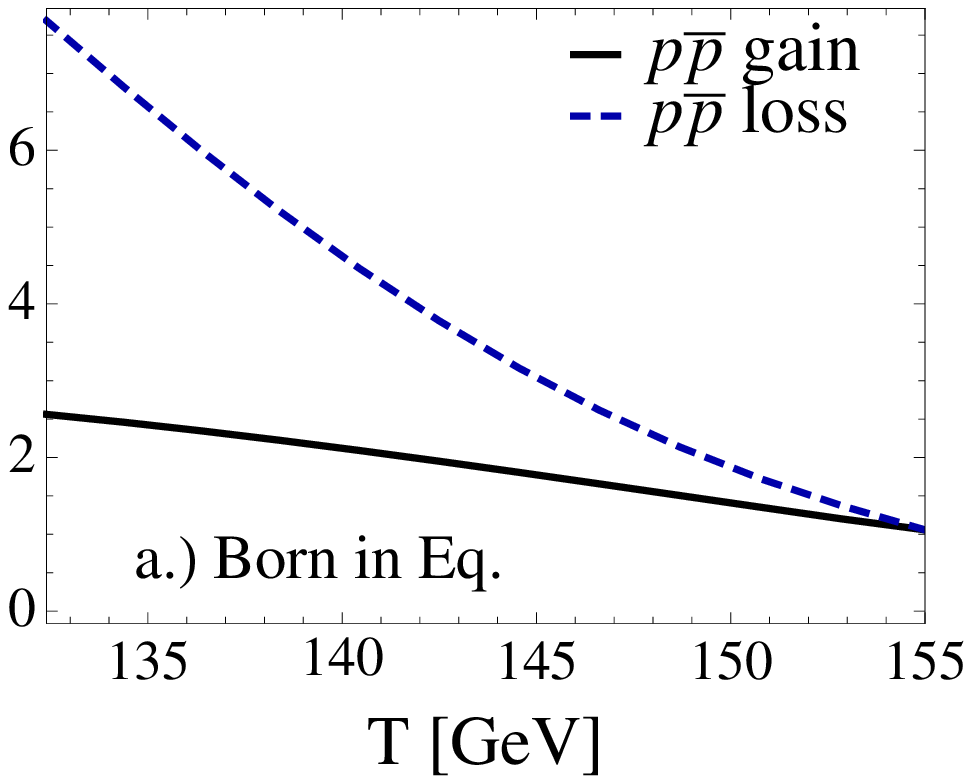} \\ 
\includegraphics[width=3.in]{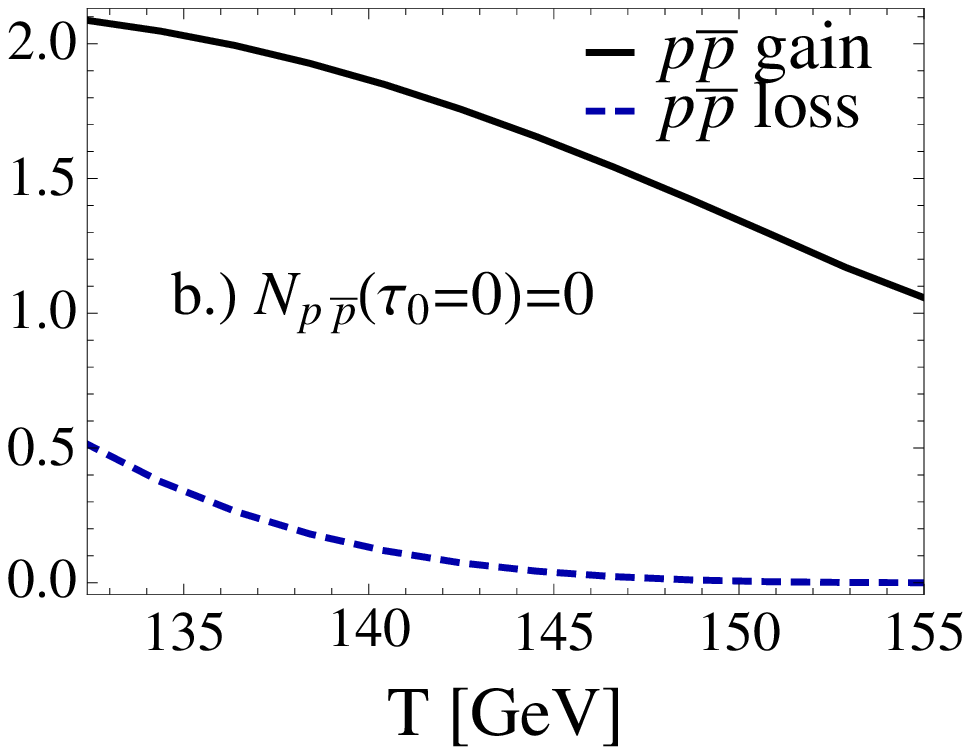}  
\caption{The total number of proton pairs $p\bar{p}$ gained (solid black line) as well as the number of $p\bar{p}$ pairs lost (blue short dashed line) for a.) $IC_1$ (all hadrons begin in chemical equilibrium) and b.) $IC_2$ (no initial protons).  For simplicity's sake we use only $\rho_1$, $T_{sw}=155$ MeV, $\tau_0=0.6$ fm/c, and $T_{end}=135$ MeV. } \label{fig:gainloss}
\end{figure}

When considering hadronic reactions it is vital to observe both the loss terms from hadrons decaying (or recombining) into other hadrons and the gain term from the reverse reactions in order to maintain detailed balance.  According
to detailed balance, the rate for the loss and gain terms must be identical in thermal equilibrium. 
Thus, neither the loss term nor the gain term  can be ignored.  In this paper we described the reactions of Hagedorn states using master equations, which naturally take into account detailed balance.  For the $X\bar{X}$ in Eq.\ (\ref{eqn:decay}) the gain and loss terms are described through:
\begin{eqnarray}\label{eqn:gainloss}
\dot{N}_{X\bar{X}}^{gain}&=&\sum_{i}\Gamma_{i,X\bar{X}} N_{i} \nonumber\\
\dot{N}_{X\bar{X}}^{loss}&=&\sum_{i}\Gamma_{i,X\bar{X}} N_{i}^{eq}\left(\frac{N_{\pi}}{N_{\pi}^{eq}}\right)^{\langle n_{i,x}\rangle} \left(\frac{N_{X\bar{X}}}{N_{X\bar{X}}^{eq}}\right)^2
\end{eqnarray}

In Fig.\ \ref{fig:gainloss} the total number of protons gained are shown in solid black lines as well as the number of proton pairs lost in blue short dashed lines.  Here we considered only $\rho_1$, $T_{sw}=155$ MeV, $\tau_0=0.6$ fm/c, and $T_{end}=133$ MeV because we were interested in the qualitative effect of the forward and back reactions, variations in the parameters would produce similar results. When all the hadrons begin in chemical equilibrium otherwise known as ''born in equilibrium" then the loss term plays a significant role in lowering the number of protons. In fact, at lower temperatures the loss term dominates and roughly 7 proton anti-proton pairs are lost for every 3 pairs created.   One can clearly see that models without backreactions would have an overpopulation of baryons.  Furthermore, even in the case where there are no initial protons the loss term plays a role. Roughly one fourth of the protons still recombine to form other hadrons over the course of the evolution of the system.  

The gain term also plays a significant role.  Without the gain term in Fig.\ \ref{fig:gainloss} b.) no proton anti-proton pairs would be produced and the gain term is also clearly needed even with the protons start in chemical equilibrium too ( a.) from Fig.\ \ref{fig:gainloss}). When the protons start in chemical equilibrium one can see in Fig.\ \ref{fig:gainloss}  a.) that still a number of $p\bar{p}$ pairs are produced.  Without the gain term in a.) the annihilation of  $p\bar{p}$ pairs would happen too quickly.  

One may wonder why the gain and loss rates are not equal in Fig.\ \ref{fig:gainloss}.  This is due to the expanding (cooling) system where the equilibrium values change over time.  If one were to hold the system at a constant temperature then the gain and loss terms would be effectively identical after a certain period of time, known as the chemical equilibration time.  Analytical estimates for this time scale were shown in \cite{NoronhaHostler:2009cf} as were calculations where the temperature was held constant. However, because our system is consistently cooling then the $N_{eq}$ values in our rate equations are monotonically decreasing, which means that the loss term will consistently be larger as the system attempts to cool towards the equilibrium values at temperature T.  However, one can clearly see that in Fig.\ \ref{fig:gainloss} b.) that the gain and loss terms are converging and the difference is small.  This is another indication that when the proton anti-proton pairs start underpopulated that they are able to reach equilibrium better. 

\section{RHIC}

\begin{figure}
\includegraphics[width=3.5in]{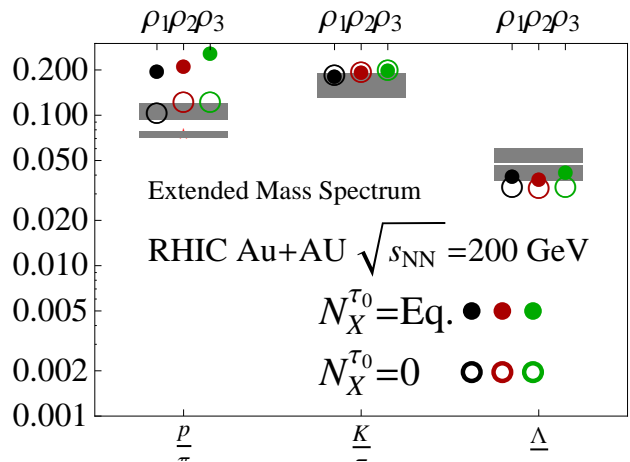}  
\caption{The total number of proton pairs $p\bar{p}$ gained (solid black line) as well as the number of $p\bar{p}$ pairs lost (blue short dashed line) for a.) $IC_1$ (all hadrons begin in chemical equilibrium) and b.) $IC_2$ (no initial protons).  For simplicity's sake we use only $\rho_1$, $T_{sw}=155$ MeV, $\tau_0=0.6$ fm/c, and $T_{end}=135$ MeV. } \label{fig:rhic}
\end{figure}

Our setup in this paper is somewhat different than in our previous works \cite{NoronhaHostler:2007jf,NoronhaHostler:2009cf} since we are concerned with emulating the general hydrodynamical modeling combined with hadronization and decays, it is natural to wonder if our model works at RHIC energies also.  

The primary difference between LHC and RHIC within our model is that we use a later initial time i.e. $\tau_0=1$ fm and the total number of pions in our system is smaller (874 at RHIC vs. 2197.5 at LHC), which effects the volume expansion in Eq.\ (\ref{eqn:temptim}).  Furthermore, the particle ratios are different.  For RHIC $Au+Au$ $\sqrt{s_{NN}}=200$ GeV the particle ratios are significantly larger than for LHC (specifically the $p/\pi$ and $\Lambda/\pi$) and are as follows: $p/\pi=0.106$, $\bar{p}/\pi= 0.082$, $K/\pi=0.156$, $\bar{K}/\pi=0.15$, $\Lambda/\pi=0.054$, and $\bar{\Lambda}/\pi=0.041$  from STAR \cite{STAR} and  $p/\pi=0.100$, $\bar{p}/\pi= 0.075$, $K/\pi=0.174 $, $\bar{K}/\pi=0.162$  from PHENIX \cite{PHENIX}. One can see that both the  $p/\pi$  and the $\Lambda/\pi$ ratios are  almost double those from LHC whereas the $K/\pi$ is only slighty higher.  

We find that while the proton to pion ratio and kaon to pion ratios are relatively easy to fit, the $\Lambda/\pi$'s tend to be underpopulated if we use the exact standard setup as we did for LHC.  However, if one increases the maximum mass up to $MM=6$ GeV one is then able to fit the experimental values for $\rho_{2}$ and $\rho_{3}$ as one can see in Fig.\ \ref{fig:rhic}.  $\rho_{1}$ is somewhat harder to fit and only manages to fit the experimental particle ratios if one increases the chemical freeze-out temperature to $T_{sw}=165$ MeV.  Alternatively, if one increases the decay width or  further increases the maximum mass one can also fit the experimental particle yields (for instance, in our previous work \cite{NoronhaHostler:2008ju,NoronhaHostler:2009cf} we used a larger maximum mass and a switching temperature identical to the Hagedorn temperature and were able to fit the RHIC particle ratios).  

This could possibly be an indication that our assumption of $\langle n_{i,p}\rangle=\langle n_{i,\Lambda}\rangle$ is inadequate (especially for $\rho_{1}$) and one would, indeed, expect that less pions should be produced in the decay $HS\leftrightarrow n\pi+\Lambda\bar{\Lambda}$ than for $HS\leftrightarrow n\pi+p\bar{p}$ simply because of the $\Lambda$'s larger mass. If one were to decrease the size of  $\langle n_{i,\Lambda}\rangle$, it would decrease the total number of pions in the system and then increase the $\Lambda/\pi$ ratio.  Conversely, it could also indicate that we need a larger maximum mass.  As we saw for LHC, we still are able to obtain a reasonable fit to the experimental particle ratios when we increase the maximum mass.  Another simple explanation could be that at RHIC the $\Lambda$ have at least a non-zero initial population or that our model assuming zero chemical potential could be further improved through a small chemical potential.  

Comparing RHIC and LHC, we find that the hydrodynamical expansion is quite a bit shorter for RHIC than LHC.  For instance, the time spent within in the hydrodynamical expansion is $\Delta \tau_{hydro}\approx3.5$, $5.5$, and $8.5$ fm for $\rho_1-\rho_3$ at RHIC, respectively, whereas $\Delta \tau_{hydro}\approx11$ fm for all $\rho$'s at LHC.  Furthermore, the expansion within the hadron gas phase lasts for $\Delta \tau_{hg}\approx4$, $6$, and $9$ fm for RHIC versus $\Delta \tau_{hg}\approx5-6$ fm for all $\rho$'s at LHC.  This implies that the hydrodynamical phase is longer for LHC whereas the hadron gas phase if roughly equivalent for both energies.  

\section{Conclusions}

In this paper, we show that our extended mass spectra model fits the experimentally measured particle ratios at ALICE $\sqrt{s}_{NN}=2.7$ TeV for certain model assumptions regarding the hadron mass spectrum.  While each type of mass spectrum description prefers a slightly different type of decays, we were not able to find a large effect coming from the mass spectra in the comparison to experimental data.  The one difference is that $\rho_3$ generally took slightly longer to reach experimental data points compared to $\rho_1$ and $\rho_2$  (roughly 1-2 fm longer that translates to having $T_{end}$ be consistently about 5 MeV lower than for $\rho_1$ and $\rho_2$).  As mentioned previously, $\rho_3$ prefers two body decays, which are more likely to last longer, thus, it is unsurprising that $T_{end}$ is lower for $\rho_3$.  While it is not clear what type of decays $\rho_1$ indicates, $\rho_2$ gives  preference to multi-hadronic decays \cite{Frautschi:1971ij}, which has not yet been adequately included into transport models while maintaining detailed balance. We do find, though, that if a longer amount of time is given for the hadron gas phase  even $\rho_3$ is able to match the experimentally measured particle ratios (or conversely larger decay widths or a larger maximum mass). This indicates that using this mass spectrum one can consistently implement the effect of Hagedorn states into hadronic transport codes.  

Due to the strong model dependence of $T_{end}$, we caution against using $T_{end}$ as a prediction for the chemical freeze-out temperature.  While on average we find $T_{end}\approx 130$ MeV, that is for a maximum mass of $MM=3$ GeV and the decay width in Eq.\ (\ref{eqn:decaywidths}).  If one increases the maximum mass or the decay width, one can easily increase the ending temperature to the range of $T_{end}=140-150$ MeV, which is the range often provided from thermal fit perdictions.  Furthermore, increasing the switching temperature also increases $T_{end}$.  Until further limitations can be set on the description of the extended mass spectrum it is difficult to use this setup to predict the chemical freeze-out temperature. 

Furthermore, we found particle ratios are sensitive to too small of decay widths in our implementation. We find that increasing the decay widths  does not affect our fits to experimental values, however, there does appear to be a limit on how low we could decrease the decay width before we are no longer able to reproduce the $K/\pi$ and $\Lambda/\pi^+$ ratio.  It may be, however, that with a smaller decay width one then requires a larger maximum mass.  

We also find that by increasing the switching temperature from $T_{sw}=155$ MeV to $T_{sw}=165$ MeV that the particle ratios do not fit the experimental data quite as well.  This fits nicely within the framework of previous work that found that the Hagedorn resonance gas was only able to fit the lattice equation of state up until $T=155$ MeV \cite{Majumder:2010ik,NoronhaHostler:2012ug} and it appears that this work also indicates that a lower switching temperature at $T_{sw}=155$ MeV or below is needed.

While previous results have suggested that final state interactions could account for the low $p/\pi$ ratio measured at ALICE \cite{Steinheimer:2012rd}, in this paper we found that the inclusion of Hagedorn states within the extended mass spectrum model can adequately reproduce this ratio when the protons (and kaons and lambda) begin underpopulated. Furthermore, even non-strange, mesonic Hagedorn states are able to quickly populate $\Lambda$ baryons, although it is likely that strange, baryonic Hagedorn states could more effectively populate $\Lambda$ baryons.  Such a scenario would involve large, heavy Hagedorn states (or multiple Hagedorn states) decaying into and combining with pions to eventually produce heavier mesons and baryons (such as lambdas, cascades, and omegas etc).   However,  when all hadrons are at chemical equilibrium at the switching temperature then the proton to pion ratio is consistently overpopulated.  In this scenario, one could allow for a longer expanding fireball (e.g. $T_{end}<100$ MeV) where one could then still obtain the $p/\pi$ and  $K/\pi$ ratios but then the $\Lambda/\pi^+$ would be vastly underpopulated. We also point out the necessity including both the gain and loss terms required within detailed balance without which it is not evident that the case of all hadrons being born in chemical equilibrium would cause such a large overpopulation of the $p/\pi$ ratio.

These findings indicate that there is a further need to investigate heavy, quickly decaying resonances within heavy-ion collisions because they may be relevant to  understanding particle ratios at the LHC. Further work is needed to include the effects of Hagedorn states into hybrid and transport models (such as the form in  \cite{Beitel:2014kza}) to realistically distinguish the contribution of highly excited, yet, hadronic states from that of the deconfined quark gluon plasma state of matter.

J.~Noronha-Hostler acknowledges
Funda\c{c}\~{a}o de Amparo \`{a} Pesquisa do Estado de S\~{a}o Paulo
(FAPESP)  for financial support. This work was supported by the Bundesministerium
f¨ur Bildung und Forschung (BMBF), the HGS-HIRe and
the Helmholtz International Center for FAIR within the
framework of the LOEWE program launched by the
State of Hesse.



\begin{thebibliography}{99}

\bibitem{thermalmodels}
  P.~Braun-Munzinger, K.~Redlich and J.~Stachel,
  arXiv:nucl-th/0304013;
  P.~Braun-Munzinger, J.~Stachel, J.~P.~Wessels and N.~Xu,
  Phys.\ Lett.\  B {\bf 344}, 43 (1995);Phys.\ Lett.\  B {\bf 365}, 1 (1996);
  J.~Cleymans, D.~Elliott, A.~Keranen and E.~Suhonen,
  Phys.\ Rev.\  C {\bf 57}, 3319 (1998);
  J.~Cleymans, H.~Oeschler and K.~Redlich,
  Phys.\ Rev.\  C {\bf 59}, 1663 (1999);
  R.~Averbeck, R.~Holzmann, V.~Metag and R.~S.~Simon,
  Phys.\ Rev.\  C {\bf 67}, 024903 (2003);
  P.~Braun-Munzinger, I.~Heppe and J.~Stachel,
  Phys.\ Lett.\  B {\bf 465}, 15 (1999).
  J.~Cleymans, H.~Satz, E.~Suhonen and D.~W.~von Oertzen,
  Phys.\ Lett.\  B {\bf 242}, 111 (1990);
  J.~Cleymans and H.~Satz,
  Z.\ Phys.\  C {\bf 57}, 135 (1993);
  F.~Becattini, M.~Gazdzicki and J.~Sollfrank,
  Eur.\ Phys.\ J.\  C {\bf 5}, 143 (1998);
  F.~Becattini, J.~Cleymans, A.~Keranen, E.~Suhonen and K.~Redlich,
  Phys.\ Rev.\  C {\bf 64}, 024901 (2001); G.~Torrieri and J.~Rafelski,
  Phys.\ Lett.\  B {\bf 509}, 239 (2001); G.~Torrieri, S.~Steinke, W.~Broniowski, W.~Florkowski, J.~Letessier and J.~Rafelski,
  Comput.\ Phys.\ Commun.\  {\bf 167}, 229 (2005);
  S.~Wheaton and J.~Cleymans,
  Comput.\ Phys.\ Commun.\  {\bf 180}, 84 (2009);


A.~Kisiel, T.~Taluc, W.~Broniowski and W.~Florkowski,
  Comput.\ Phys.\ Commun.\  {\bf 174}, 669 (2006) .
\bibitem{StatModel}
  C.~Spieles, H.~Stoecker and C.~Greiner,
  Eur.\ Phys.\ J.\  C {\bf 2}, 351 (1998)
\bibitem{Schenke:2003mj}
  B.~Schenke and C.~Greiner,
  J.\ Phys.\ G {\bf 30}, 597 (2004).

\bibitem{RHIC}
  P.~Braun-Munzinger, D.~Magestro, K.~Redlich and J.~Stachel,
  Phys.\ Lett.\  B {\bf 518}, 41 (2001);
  W.~Florkowski, W.~Broniowski and M.~Michalec,
  Acta Phys.\ Polon.\  B {\bf 33}, 761 (2002);
  W.~Broniowski and W.~Florkowski,
  Phys.\ Rev.\  C {\bf 65}, 064905 (2002);
  M.~Kaneta and N.~Xu,
  arXiv:nucl-th/0405068;
  J.~Adams {\it et al.}  [STAR Collaboration],
  Nucl.\ Phys.\  A {\bf 757}, 102 (2005).

\bibitem{Andronic:2005yp}
  A.~Andronic, P.~Braun-Munzinger and J.~Stachel,
  Nucl.\ Phys.\  A {\bf 772}, 167 (2006).
\bibitem{Manninen:2008mg}
  J.~Manninen and F.~Becattini,
  Phys.\ Rev.\  C {\bf 78}, 054901 (2008).
\bibitem{NoronhaHostler:2009tz} 
  J.~Noronha-Hostler, H.~Ahmad, J.~Noronha and C.~Greiner,
  Phys.\ Rev.\ C {\bf 82}, 024913 (2010)
  [arXiv:0906.3960 [nucl-th]].
\bibitem{freezeoutline}
  P.~Braun-Munzinger and J.~Stachel,
  Nucl.\ Phys.\  A {\bf 638}, 3 (1998);
  J.~Cleymans and K.~Redlich,
  Phys.\ Rev.\ Lett.\  {\bf 81}, 5284 (1998);Phys.\ Rev.\  C {\bf 60}, 054908 (1999);
  J.~Cleymans,
  J.\ Phys.\ G {\bf 35}, 044017 (2008);
  J.~Cleymans, R.~Sahoo, D.~K.~Srivastava and S.~Wheaton,
  Eur.\ Phys.\ J.\ ST {\bf 155}, 13 (2008)
  J.~Cleymans, R.~Sahoo, D.~P.~Mahapatra, D.~K.~Srivastava and S.~Wheaton,
  Phys.\ Lett.\  B {\bf 660}, 172 (2008);
  J.~Cleymans, H.~Oeschler, K.~Redlich and S.~Wheaton,
  J.\ Phys.\ G {\bf 32}, S165 (2006).


\bibitem{Abelev:2012wca} 
  B.~Abelev {\it et al.}  [ALICE Collaboration],
  Phys.\ Rev.\ Lett.\  {\bf 109}, 252301 (2012)
  [arXiv:1208.1974 [hep-ex]].
\bibitem{Abelev:2013xaa} 
  B.~B.~Abelev {\it et al.}  [ALICE Collaboration],
  Phys.\ Rev.\ Lett.\  {\bf 111}, 222301 (2013)
  [arXiv:1307.5530 [nucl-ex]].
  

\bibitem{LHC}
  N.~Armesto {\it et al.},
  J.\ Phys.\ G {\bf 35}, 054001 (2008).
 
\bibitem{Steinheimer:2012rd} 
  J.~Steinheimer, J.~Aichelin and M.~Bleicher,
  Phys.\ Rev.\ Lett.\  {\bf 110}, 042501 (2013)
  [arXiv:1203.5302 [nucl-th]].
\bibitem{Petersen:2008dd}  
H.~Petersen, J.~Steinheimer, G.~Burau, M.~Bleicher and H.~Stocker,
  Phys.\ Rev.\ C {\bf 78}, 044901 (2008)
  [arXiv:0806.1695 [nucl-th]].
\bibitem{urqmd}
S.~A.~Bass, M.~Belkacem, M.~Bleicher, M.~Brandstetter, L.~Bravina, C.~Ernst, L.~Gerland and M.~Hofmann {\it et al.},
  Prog.\ Part.\ Nucl.\ Phys.\  {\bf 41}, 255 (1998)
  [Prog.\ Part.\ Nucl.\ Phys.\  {\bf 41}, 225 (1998)]
  [nucl-th/9803035].
M.~Bleicher, E.~Zabrodin, C.~Spieles, S.~A.~Bass, C.~Ernst, S.~Soff, L.~Bravina and M.~Belkacem {\it et al.},
  J.\ Phys.\ G {\bf 25}, 1859 (1999)
  [hep-ph/9909407].
\bibitem{Ratti:2011au} 
  C.~Ratti, R.~Bellwied, M.~Cristoforetti and M.~Barbaro,
  Phys.\ Rev.\ D {\bf 85}, 014004 (2012)
  [arXiv:1109.6243 [hep-ph]].

  
\bibitem{decaysSPS} 
  R.~Rapp and E.~V.~Shuryak,
  Phys.\ Rev.\ Lett.\  {\bf 86}, 2980 (2001)
  [hep-ph/0008326].
  C.~Greiner,
  AIP Conf.\ Proc.\  {\bf 644}, 337 (2003)
  [nucl-th/0208080].
  C.~Greiner,
  Heavy Ion Phys.\  {\bf 14}, 149 (2001)
  [nucl-th/0011026].
  C.~Greiner and S.~Leupold,
  J.\ Phys.\ G {\bf 27}, L95 (2001)
  [nucl-th/0009036].
\bibitem{decaysRHIC} 
P.~Braun-Munzinger, J.~Stachel and C.~Wetterich,
  Phys.\ Lett.\ B {\bf 596}, 61 (2004)
  .
   S.~Pal and W.~Greiner,
  Phys.\ Rev.\ C {\bf 87}, no. 5, 054905 (2013).
\bibitem{Greiner:2004vm}
  C.~Greiner {\it et al.}  
  J.\ Phys.\ G {\bf 31}, S725 (2005).
\bibitem{NoronhaHostler:2007jf} 
  J.~Noronha-Hostler, C.~Greiner and I.~A.~Shovkovy,
  Phys.\ Rev.\ Lett.\  {\bf 100}, 252301 (2008)
  [arXiv:0711.0930 [nucl-th]].
\bibitem{NoronhaHostler:2009cf} 
  J.~Noronha-Hostler, M.~Beitel, C.~Greiner and I.~Shovkovy,
  Phys.\ Rev.\ C {\bf 81}, 054909 (2010)
  [arXiv:0909.2908 [nucl-th]].
\bibitem{Hagedorn:1965st} 
  R.~Hagedorn,
  Nuovo Cim.\ Suppl.\  {\bf 3}, 147 (1965);
R.~Hagedorn,
  Nuovo Cim.\ A {\bf 56}, 1027 (1968).
\bibitem{Broniowski:2004yh} 
  W.~Broniowski, W.~Florkowski and L.~Y.~.Glozman,
  Phys.\ Rev.\ D {\bf 70}, 117503 (2004)
  [hep-ph/0407290].
\bibitem{NoronhaHostler:2008ju} 
  J.~Noronha-Hostler, J.~Noronha and C.~Greiner,
  Phys.\ Rev.\ Lett.\  {\bf 103}, 172302 (2009)
  [arXiv:0811.1571 [nucl-th]].
\bibitem{NoronhaHostler:2012ug} 
  J.~Noronha-Hostler, J.~Noronha and C.~Greiner,
  Phys.\ Rev.\ C {\bf 86}, 024913 (2012)
  [arXiv:1206.5138 [nucl-th]].
  
  \bibitem{Pal:2010es} 
  S.~Pal,
  Phys.\ Lett.\ B {\bf 684}, 211 (2010)
 .
\bibitem{Borsanyi:2010cj} 
  S.~Borsanyi, G.~Endrodi, Z.~Fodor, A.~Jakovac, S.~D.~Katz, S.~Krieg, C.~Ratti and K.~K.~Szabo,
  JHEP {\bf 1011}, 077 (2010)
  [arXiv:1007.2580 [hep-lat]].
\bibitem{Majumder:2010ik} 
  A.~Majumder and B.~Muller,
  Phys.\ Rev.\ Lett.\  {\bf 105}, 252002 (2010)
  [arXiv:1008.1747 [hep-ph]].
\bibitem{Noronha-Hostler:2013ria} 
  J.~Noronha-Hostler, J.~Noronha, G.~S.~Denicol, R.~P.~G.~Andrade, F.~Grassi and C.~Greiner,
  arXiv:1302.7038 [nucl-th].
  
\bibitem{Moretto:2005iz} 
  L.~G.~Moretto, K.~A.~Bugaev, J.~B.~Elliott and L.~Phair,
  Europhys.\ Lett.\  {\bf 76}, 402 (2006)
  [nucl-th/0504010].

\bibitem{Begun:2009an} 
  V.~V.~Begun, M.~I.~Gorenstein and W.~Greiner,
  J.\ Phys.\ G {\bf 36}, 095005 (2009)
  [arXiv:0906.3205 [nucl-th]].
  
  Zakout:2006zj,Zakout:2007nb,Ferroni:2008ej,Bugaev:2008iu,Ivanytskyi:2012yx
  
\bibitem{Zakout:2006zj} 
  I.~Zakout, C.~Greiner and J.~Schaffner-Bielich,
  Nucl.\ Phys.\ A {\bf 781}, 150 (2007)
  [nucl-th/0605052].
\bibitem{Zakout:2007nb} 
  I.~Zakout and C.~Greiner,
  Phys.\ Rev.\ C {\bf 78}, 034916 (2008)
  [arXiv:0709.0144 [nucl-th]].
\bibitem{Ferroni:2008ej} 
  L.~Ferroni and V.~Koch,
  Phys.\ Rev.\ C {\bf 79}, 034905 (2009)
  [arXiv:0812.1044 [nucl-th]].
\bibitem{Bugaev:2008iu} 
  K.~A.~Bugaev, V.~K.~Petrov and G.~M.~Zinovjev,
  Phys.\ Rev.\ C {\bf 79}, 054913 (2009)
  [arXiv:0807.2391 [hep-ph]].
\bibitem{Ivanytskyi:2012yx} 
  A.~I.~Ivanytskyi, K.~A.~Bugaev, A.~S.~Sorin and G.~M.~Zinovjev,
  Phys.\ Rev.\ E {\bf 86}, 061107 (2012)
  [arXiv:1211.3815 [nucl-th]].
  
\bibitem{Bazavov:2014xya} 
  A.~Bazavov, H.~-T.~Ding, P.~Hegde, O.~Kaczmarek, F.~Karsch, E.~Laermann, Y.~Maezawa and S.~Mukherjee {\it et al.},
  arXiv:1404.6511 [hep-lat].

\bibitem{Frautschi:1971ij} 
  S.~C.~Frautschi,
  Phys.\ Rev.\ D {\bf 3}, 2821 (1971).
\bibitem{Greiner:1993jn} 
  C.~Greiner, C.~Gong and B.~Muller,
  Phys.\ Lett.\ B {\bf 316}, 226 (1993).
  
\bibitem{Abelev:2013vea} 
  B.~Abelev {\it et al.}  [ALICE Collaboration],
  Phys.\ Rev.\ C {\bf 88}, 044910 (2013)
  
\bibitem{Lizzi:1990na} 
  F.~Lizzi and I.~Senda,
  Phys.\ Lett.\ B {\bf 244}, 27 (1990).
\bibitem{Lizzi:1990if} 
  F.~Lizzi and I.~Senda,
  Nucl.\ Phys.\ B {\bf 359}, 441 (1991).
  
  
\bibitem{Liu}
  F.~M.~Liu, K.~Werner and J.~Aichelin,
  Phys.\ Rev.\ C {\bf 68} (2003) 024905;
  F.~M.~Liu, et. al.,
  J.\ Phys.\ G {\bf 30} (2004) S589;
  Phys.\ Rev.\ C {\bf 69} (2004) 054002.



\bibitem{Pal:2005rb} 
  S.~Pal and P.~Danielewicz,
  Phys.\ Lett.\ B {\bf 627}, 55 (2005)
  [nucl-th/0505049].

  
   \bibitem{Beitel:2014kza} 
  M.~Beitel, K.~Gallmeister and C.~Greiner,
  arXiv:1402.1458 [hep-ph].

\bibitem{STAR} 
  O.~Y.~.Barannikova [STAR Collaboration],
  nucl-ex/0403014.

  J.~Adams {\it et al.}  [STAR Collaboration],
  Nucl.\ Phys.\ A {\bf 757}, 102 (2005)
  [nucl-ex/0501009].
\bibitem{PHENIX} 
  S.~S.~Adler {\it et al.}  [PHENIX Collaboration],
  Phys.\ Rev.\ C {\bf 69}, 034909 (2004)
  [nucl-ex/0307022].
\end{thebibliography}
\end{document}